\documentclass[12pt]{iopart}

\usepackage{iopams}

\usepackage{graphicx}% Include figure files
\usepackage{dcolumn}% Align table columns on decimal point
\usepackage{bm}% bold math

\newcommand{\Ji}{\mathit J_{\mathrm{sd}}}
\newcommand{\rv}{\bi r}

\begin{document}

\title{Microscopic theory of the Coulomb based exchange coupling in magnetic tunnel junctions}

\author{O.~G.~Udalov}
\address{Department of Physics and Astronomy, California State University Northridge, Northridge, CA 91330, USA}
\address{Institute for Physics of Microstructures, Russian Academy of Science, Nizhny Novgorod, 603950, Russia}
\ead{oleg.udalov@csun.edu}

\author{I.~S.~Beloborodov}
\address{Department of Physics and Astronomy, California State University Northridge, Northridge, CA 91330, USA}

\pacs{75.50.Tt 75.75.Lf	75.30.Et 75.75.-c}

\date{\today}

\begin{abstract}
We study interlayer exchange coupling based on the many-body Coulomb interaction
between conduction electrons in magnetic tunnel junction.
This mechanism complements the known interaction between magnetic layers 
based on virtual electron hopping (or spin currents). We find that these two mechanisms have
different behavior on system parameters. The Coulomb based coupling may exceed the hopping based exchange.
We show that the Coulomb based exchange interaction, in contrast to the hopping based coupling, 
depends strongly on the dielectric constant of the insulating layer.
The dependence of the interlayer exchange interaction on the dielectric 
properties of the insulating layer in magnetic tunnel junction is similar
to magneto-electric effect where electric and magnetic degrees of freedom are coupled.
We calculate the interlayer coupling as a function of temperature and 
electric field for magnetic tunnel junction with
ferroelectric layer and show that the exchange interaction between magnetic leads 
has a sharp decrease in the vicinity of the ferroelectric phase transition
and varies strongly with external electric field.
\end{abstract}

\submitto{\JPCM}

\noindent{\it magneto-electric effect, multiferroics, Coulomb interaction, magnetic tunnel junction, exchange interaction}:

\maketitle

\section{Introduction}

Magnetic tunnel junctions (MTJ) are of great importance these days
because of their promise for next generation memory cells~\cite{Worledge2015,Dau2007,GUPTA2004,Prinz1998,Cho2002,Yoda2014}.
However, there are several issues to create such memory cells.
One of them is control of magnetic state of ferromagnetic (FM) leads in MTJ.
This problem requires a fundamental research and understanding of spin currents
and interlayer exchange coupling (IEC) mechanisms in MTJ.
The IEC was studied theoretically~\cite{Julliere1975,Slonczewski1989,Bruno1995,
Vedyayev2005,Vega2001,Tsymbal2006,Stiles2009,Yuasa2007,Tsymbal2010} and experimentally~\cite{Grunberg2002,Siegel2003,Tsymbal2006,Grunberg2002,Schuhl2002,Jonge1997,
Kita2007,Dijken2010,Korecki2014}
in numerous papers. These works show that there are several phenomena responsible for
magnetic coupling between FM layers in MTJ. A weak FM coupling appears due to magneto-dipole (MD)
interaction between correlated roughness in MTJ interfaces (orange peel effect)~\cite{Tanner1999,Kools1995}.
Spin currents across the insulating barrier in MTJ produce the hopping
based exchange interaction~\cite{Julliere1975,Slonczewski1989, Bruno1995} leading to the FM or
antiferromagnetic (AFM) coupling depending on the system parameters~\cite{Siegel2003,Schuhl2002}.
In addition, the IEC was studied for insulating layer with impurities~\cite{Vedyayev2005, Tsymbal2006imp,Tsymbal2006}.
This mechanism gives the AFM contribution to the total magnetic coupling between FM leads.
In this paper we propose a different mechanism of IEC in MTJ based on many-body Coulomb interaction.

The hopping based IEC in MTJ was studied both
analytically~\cite{Julliere1975,Slonczewski1989, Bruno1995,Vedyayev2005,Stiles2009,Tsymbal2010} and using
the \textit{ab initio} calculations~\cite{Vega2001,Tsymbal2006,Yuasa2007}.
To derive the IEC the single particle Hamiltonian was used
with kinetic term and the spin-dependent single-particle potential. The
many-body interaction was not taken into account when calculating the IEC.

The exchange coupling exists due to many-particle effects,
for example due to the Coulomb interaction.
The indirect Coulomb interaction (exchange) between electrons localized at different
atoms is responsible for FM and AFM states in magnetic insulators~\cite{Vons}.
The indirect Coulomb interaction between conduction and localized electrons in FM metals
results in spin subband splitting of the conduction band~\cite{Vons}.
In this paper we study the IEC based on the inter-electron Coulomb interaction between
two FM metallic leads separated by the insulating layer.

The basic idea behind this mechanism is related to the fact that the wave functions of electrons located
at different FM leads are overlapped inside the insulating layer. Since the screening effects inside the
insulator are weak the long-range Coulomb interaction between these electrons exists and it has
the spin-dependent part (exchange interaction)~\cite{Landau3}. Here we study the
IEC due to this indirect spin-dependent part of the Coulomb interaction and
show that this contribution is comparable and even bigger than the hopping
based exchange coupling.

The most important feature of the Coulomb based exchange interaction is its
dependence on the dielectric permittivity of the insulating layer. This dependence
can be used to distinguish this contribution from other contributions to the exchange coupling.
Also this peculiarity allows to realize the magneto-electric (ME) coupling~\cite{Scott2006,Spal2007}
in MTJ with ferroelectric (FE) insulating barrier. 
Generally, ME effects in MTJ are known even without the FE barrier. 
Applying voltage to the MTJ leads to the deformation of the barrier potential profile 
and to variation of hopping based IEC~\cite{Brataas2008,Lee2009,SUZUKI2008,Ralph2008}]. Spin transfer torque effect also appear in the voltage biased MTJ  leading to magnetization dynamics~\cite{Slonczewski1989,Slonczewski2005,Butler2006,Brataas2008,Lee2009,SUZUKI2008,Ralph2008}. 
The ME effect in MTJ with FE barrier was studied in several works~\cite{Vedyayev2005}. 
The FE creates the surface charges at the interfaces between the barrier and FM leads. 
These charges modify the potential barrier profile and influence the IEC. 
Switching FE polarization with electric field allows to change the magnitude of IEC. 
All mentioned effects are related to the hopping based IEC. In the current manuscript 
we propose a completely different mechanism based on many-body effects.

The paper is organized as follows. In section~\ref{Sec:model} we introduce
the model for MTJ. In sections~\ref{Sec:WF} and \ref{Coulomb} we calculate the inter-electron Coulomb interaction
and the IEC in the system. We discuss and compare the Coulomb and the
hopping based IEC in section.~\ref{Sec:Discuss}. We consider ME coupling in
MTJ with FE barrier in section~\ref{Sec:Discuss}. Finally, we discuss
validity of our theory in section~\ref{Sec:Val}.

\section{The model}\label{Sec:model}

\begin{figure}
\includegraphics[width=0.6\columnwidth]{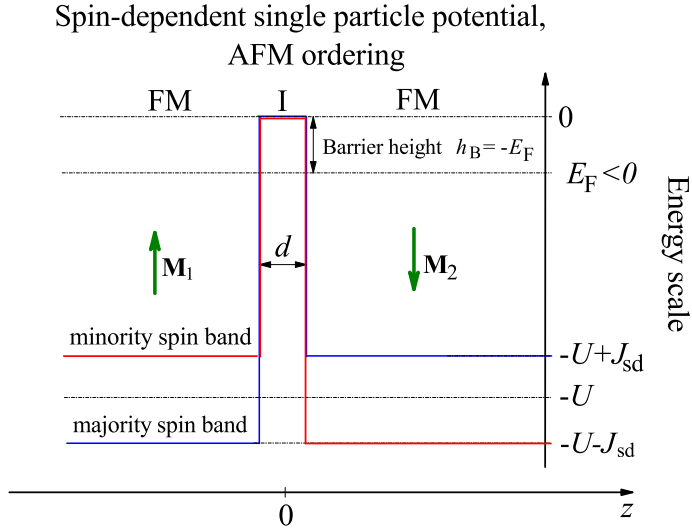}
\caption{(Color online) Schematic picture of potential energy profiles
for electron with spin ``up'' and ``down'' states for AFM configuration of
leads magnetic moments $\bi M_{1,2}$. Zero energy corresponds to the top of energy barrier for electrons in the insulator. Symbols FM and I stand for FM metal and insulator, respectively. All other notations are defined in the text. } \label{Fig:EnExp}
\end{figure}

We consider two identical FM leads separated by the insulating layer of thickness $d$ (figure~\ref{Fig:EnExp}).
The FM leads have thickness much larger than all other characteristic length scales
in the problem. We find magnetic interaction between the leads assuming that s electrons are responsible for
magnetic interaction. We assume that magneto-dipole interaction between the leads is absent.
The Hamiltonian describing delocalized electrons in the system has the form
\begin{equation}\label{Eq:HamIn}
\eqalign{
&\hat H=\hat H_0 +\hat H_{\mathrm{C}},\cr
&\hat H_0=\sum_i(\hat W_\mathrm k(\bi{r}_i)+\hat U_1(\bi{r}_i)+\hat U_2(\bi{r}_i)+\hat H_{\mathrm{1m}}(\bi{r}_i)+\hat H_{\mathrm{2m}}(\bi{r}_i)),
}
\end{equation}
where  $\hat H_\mathrm{0}$ is the single particle Hamiltonian. The term $\hat W_\mathrm k$ describes the kinetic energy of an electron. Single particle potential energy $\hat U_1=-U\Theta(-z-d/2)$ and $\hat U_2=-U\Theta(z-d/2)$, where the step function $\Theta(z)=1$, if $z>0$ and $\Theta(z)=0$, if $z<0$. We choose the zero energy
level at the top of insulating barrier (see figure~\ref{Fig:EnExp}). $\hat H_{1,2\mathrm{m}}$ describes the spin subband splitting of delocalized electrons. In the frame of Vonsovskii s-d model such a splitting appears due to exchange interaction of delocalized s electrons with localized d electrons in FM metals~\cite{Vons}. The s-d interaction couples d and s electrons in the same lead and does not produce coupling between the leads. We consider only
FM and AFM collinear configurations of the leads magnetizations. Therefore, we have
$\hat H_{1,2\mathrm{m}}=-J_\mathrm{sd}\hat \sigma_z M_{1,2}\Theta(\mp z-d/2)$ ( sign ``$-$'' stands
for the left and ``$+$'' for the right leads, respectively), with $M_1=1$ and $M_{2} = \pm 1$ for FM (``+'') and AFM (``-'') configurations.
Note that the spin subband splitting may also appear due to the exchange interaction of s electrons within a lead (see \cite{Glazman2002}). This coupling renormalizes the s-d coupling constant $J_\mathrm{sd}$ but
also does not lead to the IEC. $\hat H_{\mathrm{C}}$ is the Coulomb interaction between conduction s electrons;

In ~\cite{Slonczewski1989, Bruno1995} it was demonstrated that the magnetic exchange interaction between FM leads
may appear  due to the single particle Hamiltonian $\hat H_0$.
The magnetic interaction was considered as a result of spin currents
flowing between the leads \cite{Slonczewski1989} or as a consequence of
varying of electron energy levels density due to interference effects in the insulating layer between the leads \cite{Bruno1995}.
Using perturbation theory we consider the magnetic interaction as a result of virtual electron hopping between
the leads \cite{Bel2016ExGr}. The interlead coupling in Hamiltonian $\hat H_0$ appears as
a combination of s-d interaction and the kinetic energy term. The s-d interaction produces the
spin polarized s electrons in each lead while the kinetic energy term couples
these spin polarized electron gases in the left and the right leads.

It is known that the exchange coupling also exists due to many particle effects.
Spin dependent part of the Coulomb interaction couples magnetic moments of two electrons located at different sites
with overlapping wave functions (so called Heitler-London model)~\cite{Landau3}.
Originally this coupling was called the exchange interaction. In this manuscript we consider
the spin dependent indirect Coulomb interaction between s electrons located in different layers. 
Generally, we will sum the interaction of Heitler-London type over all pair of electrons located
in different magnetic leads.
It leads to the magnetic coupling between the FM layers. This is in contrast to
the s-d interaction where d-electrons act on s-electrons in the same layer and do not
affect directly the s-electrons in the other lead.

\subsection{General procedure of calculation of IEC}

We introduce single particle Hamiltonians $\hat H_{1,2}$ describing a single lead with
infinite insulator
\begin{equation}\label{Eq:HamSep}
\hat H_{1,2}=\hat W_{\mathrm k}+\hat U_{1,2}+\hat H_{1,2\mathrm{m}}.
\end{equation}
The eigenfunctions of the Hamiltonians are $\psi_i^s$ and $\phi_i^s$ for leads (1) and (2), respectively. The subscript $i$ stands for orbital state and the superscript $s$ denotes the spin state in the local spin coordinate system related to magnetization in the
corresponding lead. The wave functions are symmetric due to symmetry of the problem $\psi^s_i(x,y,z)=\phi^s_i(x,y,-z)$. Energies of these states are
$\epsilon_{1i}^s$ and $\epsilon_{2i}^s$. For identical leads the energies
are equal, $\epsilon_{1i}^s=\epsilon_{2i}^s=\epsilon_i^s$.
The creation and annihilation operators in the lead (1) and (2) are $\hat a^{s+}_i$ and $\hat a^{s}_i$,
and $\hat b^{s+}_i$ and $\hat b^{s}_i$, respectively.

We introduce the zero-order many-particle wave
functions $\Psi^\mathrm{AFM}_0$ and  $\Psi^\mathrm{FM}_0$ for
AFM and FM configurations of leads magnetic moments $\bi{M}_{1,2}$
(see wave functions in \ref{App:Validity}).
These wave functions describe the non-interacting FM layers ($d\to\infty$).
All states $\psi^s_i$ and $\phi^s_j$ with
energies $\epsilon^s_i<E_\mathrm F$ are filled and all states above $E_\mathrm F$ are empty.

Below we calculate the surface exchange interaction between
the FM leads, $H^\mathrm{ex}=(E^\mathrm{AFM}-E^\mathrm{FM})/S$,
where $E^{\mathrm{AFM}}$ ($E^{\mathrm{FM}}$) is the energy
of AFM (FM) state and $S$ is the surface area of the leads.
We will assume that the insulator between the leads is thick enough and
the interlayer coupling is weak and can be considered within the perturbation theory.
The exchange interaction has two contributions:
1) the hopping based exchange interaction, $H^\mathrm{ex}_\mathrm{h}$ - due to the Hamiltonian $\hat H_0$
and 2) the Coulomb based exchange interaction, $H^\mathrm{ex}_\mathrm C$ - due to the Hamiltonian $\hat H_\mathrm C$.
The first contribution was calculated in ~\cite{Bruno1995}.   The second contribution
is considered in this paper. When calculating the Coulomb based contribution we neglect corrections to the ground state $\Psi_0$ due to electron tunnelling between the leads. Validity of this approximation is discussed in the \ref{App:Validity}.
Finally, we use the first order perturbation theory to
calculate the indirect Coulomb interaction
\begin{equation}\label{Eq:ExGen}
H^\mathrm{ex}_\mathrm C=\frac{1}{S}(\langle\Psi^\mathrm{AFM}_0|\hat H_\mathrm C|\Psi^\mathrm{AFM}_0\rangle-\langle\Psi^\mathrm{FM}_0|\hat H_\mathrm C|\Psi^\mathrm{FM}_0\rangle).
\end{equation}
We consider ~(\ref{Eq:ExGen}) in the next sections.

\subsection{Estimation of the Hoping based IEC}

The hopping based exchange interaction due to Hamiltonian $\hat H_0$ can be estimated as follows.
Electron wave functions decay exponentially inside the insulator. Therefore the hopping
matrix elements between the leads depend exponentially on distance $d$ as $\rme^{-\varkappa_0 d}$,
where $\varkappa_0=\sqrt{-2m_\mathrm eE_\mathrm F/\hbar^2}$.
These matrix elements are small. The hopping based exchange interaction appears in the
second order perturbation theory \cite{Bel2016ExGr} producing the factor $\rme^{-2\varkappa_0 d}$.
Since the hopping based interlead exchange interaction is due to kinetic energy it has
the factor $\hbar^2\varkappa_0^2/(2m_e)$, which always enter the tunnelling matrix elements.
In addition, the magnetic coupling between the leads is absent
if electron gas inside the leads is not spin polarized (has no magnetic moment).
Therefore, the interlayer coupling is proportional to the spin polarization
of s-electrons, $(J_\mathrm{sd}/(U+E_\mathrm F))^2$. Finally,
the hopping based IEC can be estimated as follows, $H^\mathrm{ex}_\mathrm{h}\sim (\hbar^2\varkappa_0^2/2m_\mathrm e)(J_\mathrm{sd}/(U+E_\mathrm F))^2\rme^{-2\varkappa_0 d}$, in a qualitative agreement with \cite{Bruno1995}.

\subsection{Estimation of the Coulomb based IEC}

Similarly we estimate the s-s Coulomb based exchange interaction between the leads.
The matrix element of the indirect Coulomb interaction contains four electron wave
functions and therefore is proportional to the square of electron wave functions overlap.
The Coulomb based exchange interaction appears in the first order perturbation theory. Therefore, it
is also proportional to $\rme^{-2\varkappa_0 d}$. Similar to the hopping based coupling the Coulomb based coupling
exists only if electron gas in each lead is spin-polarized producing the
factor $(J_\mathrm{sd}/(U+E_\mathrm F))^2$. In addition, the Coulomb interaction between
two electrons inside the barrier can be estimated as $e^2/(\varepsilon d)$,
where $\varepsilon$ is the insulator dielectric constant. Combining all these factors
the Coulomb based exchange coupling can be estimated as follows,
$H^\mathrm{ex}_\mathrm{C}\sim (e^2/\varepsilon d)(J_\mathrm{sd}/(U+E_\mathrm F))^2\rme^{-2\varkappa_0 d}$.

For interlayer distance $d=1$ nm, barrier height $h_\mathrm B=-E_\mathrm F=0.2$ eV, and $\varepsilon=5$
the hopping and the Coulomb based exchange contributions are comparable.
Therefore the Coulomb based exchange coupling is important
and should be taken into account in this problem.

\section{Main results}

Our main findings are the following:

1) We calculate the exchange coupling in MTJ due to Coulomb interaction
between s electrons located at different leads. We show that the Coulomb
based IEC can exceed the hopping based exchange contribution studied by
Slonczewski, Bruno and Jullier in the past. We find that these two mechanisms
have essentially different dependence on system parameters (see figures~\ref{Fig:JvsEf1}, \ref{Fig:JvsEf2}, \ref{Fig:Jvsd}, \ref{Fig:JvsJ1}, \ref{Fig:ExTot}).

2) We find that the hopping and the Coulomb based exchange interaction
have the same small factor due to weak wave functions overlap, $\rme^{-2\varkappa_0 d}$.
In addition, both contributions have similar dependence on the s-d coupling constant.
For small spin subband splitting both contributions are proportional to $(J_\mathrm{sd}/(U+E_\mathrm F))^2$.
The hopping based IEC is related to the kinetic energy term in the system Hamiltonian
and therefore has the factor $\hbar^2 \varkappa_0^2/(2m_\mathrm e)$.
The Coulomb based exchange contribution is due to electric forces
acting between s-electrons, as a result it has the factor $e^2/(\varepsilon d)$.

3) We find the approximate analytical expressions for the Coulomb
($H^\mathrm{ex}_\mathrm{C}$) and the hopping
($H^\mathrm{ex}_\mathrm{h}$) based IEC per unit area in the limit
$J_\mathrm{sd}\ll (U+E_\mathrm F)$, $\varkappa_0\ll k_\mathrm F$ ($k_\mathrm F=\sqrt{2m_\mathrm e(U+E_\mathrm F)/\hbar^2}$)
\begin{equation}\label{Eq:ExHopApp1}
H^\mathrm{ex}_\mathrm{h}\approx -\frac{2\hbar^2}{\pi^2 m_\mathrm e d^2} \,
\frac{J^2_\mathrm{sd}}{(U+E_\mathrm F)^2} \,
\frac{\varkappa_0^5}{k^3_\mathrm F} \, \rme^{-2\varkappa_0 d},
\end{equation}
\begin{equation}\label{Eq:ExCoulApp1}
H^\mathrm{ex}_\mathrm{C}\approx \frac{\zeta e^2}{4\pi\varepsilon_0\varepsilon d} \,
\frac{J^2_\mathrm{sd}}{(U+E_\mathrm F)^2} \,
\frac{k_\mathrm F}{d} \, \rme^{-2\varkappa_0 d},
\end{equation}
where $\zeta\approx 1/150$.

4) We show that in contrast to hopping based exchange interaction,
the Coulomb based exchange is inversely proportional to the dielectric
constant $\varepsilon$ of the insulating layer (see figure~\ref{Fig:JvsEps}).

5) We calculate the IEC as a function of temperature and electric field (or voltage across the MTJ)
for MTJ with insulating layer made of tetrathiafulvalene p-chloranil complex (TTF-CA) 
and Hf$_{0.5}$Zr$_{0.5}$O$_2$ (see figures~\ref{Fig:JvsT1}, \ref{Fig:JvsE1}).
We show that in the vicinity of the FE phase transition of TTF-CA the IEC has large
variations. We find that even the FM-AFM transition may occur in MTJ for some system parameters.

6) We find that IEC as a function of electric field (or applied voltage)
shows strong variations. The electric field
can cause the transition from AFM to FM coupling: for zero field the coupling is AFM
while for finite field it is FM. This effect demonstrates the ME coupling in MTJ.

\section{Basic wave functions}\label{Sec:WF}

We use the coordinate system with $z$-axis being perpendicular to the
leads surfaces and $x$ and $y$ being perpendicular to $z$, $r_\perp=\sqrt{x^2+y^2}$.
In the absence of spin-orbit interaction the spin and the spatial parts
of wave functions are decoupled. The spin parts are $(1~0)^T$ and $(0~1)^T$
for the spin up and spin down states, respectively. The spatial part of electron wave functions
inside the leads consists of two plane waves: one going toward the leads surface and the other
being reflected from the FM/insulator interface
\begin{equation}\label{Eq:WaveFunc1}
\eqalign{
&\psi^s_i(z,\bi{r}_\perp)=\frac{\rme^{\rmi k_{z}\left(\frac{d}{2}+ z\right)}+\xi^s_\bi k \rme^{-\rmi k_z\left(\frac{d}{2}+ z\right)}}{\sqrt{\Omega}}\rme^{\rmi \bi k_\perp \bi r_\perp},~z<-d/2\cr
&\phi^s_i(z,\bi{r}_\perp)=\frac{\rme^{\rmi k_{z}\left(\frac{d}{2}- z\right)}+\xi^s_\bi k \rme^{-\rmi k_z\left(\frac{d}{2}- z\right)}}{\sqrt{\Omega}}\rme^{\rmi \bi k_\perp \bi r_\perp},~z>d/2,
}
\end{equation}
where $\xi^s_\bi k = \frac{k_z-\rmi\varkappa^s_\bi k}{k_z+\rmi\varkappa^s_\bi k}$ is the amplitude of the reflected electron wave, with $\varkappa^s_\bi k=\sqrt{2m_\mathrm e (U-s\Ji)/\hbar^2-k_z^2}$. Superscript $i$ denotes the
full set of quantum numbers ($k_x,k_y,k_z$), $\bi k_{\perp}=(k_x,k_y,0)$ and $\bi r_\perp=(x,y,0)$ and
$\Omega$ is the volume of each lead.

In the region $z>-d/2$ we have for the wave function of the left lead
\begin{equation}\label{Eq:WaveFunc2}
\psi^s_i(z,r_\perp)=\frac{\tau^s_\bi k}{\sqrt{\Omega}}\rme^{-\varkappa^s_\bi k\left(\frac{d}{2}+ z\right)}\rme^{\rmi\bi k_\perp \bi r_\perp},~z>-d/2,
\end{equation}
where $\tau^s_\bi k =\frac{2k_z}{k_z+\rmi\varkappa^s_\bi k}$ is the amplitude of the transmitted electron wave.
The electron wave function for the states in the right lead for $z<d/2$ is
\begin{equation}\label{Eq:WaveFunc3}
\phi^s_i(z,r_\perp)=\frac{\tau^s_\bi k}{\sqrt{\Omega}}\rme^{-\varkappa^s_\bi k\left(\frac{d}{2}- z\right)}\rme^{\rmi\bi k_\perp \bi r_\perp},~z<d/2.
\end{equation}
Note that $\psi$ and $\phi$ are the wave functions of isolated leads ($d\to \infty$).
Therefore, these wave functions decay exponentially in the region $z>-d/2$ for the left lead states and $z<d/2$ for the right lead states. This is in contrast to Bruno (or Slonczewki) model for hopping based IEC, where
the wave functions of states in the left lead have oscillating part in the right
lead and vice versa. In our consideration we neglect tunnelling
when calculating the Coulomb based IEC. We discuss this point in more
details in  \ref{App:Validity}.

\section{Exchange due to Coulomb interaction} \label{Coulomb}

Both terms in  ~(\ref{Eq:ExGen}) have contributions from direct and indirect Coulomb interaction.
The direct Coulomb interaction does not lead to the spin-dependent correction to the system energy and
will be omitted below. Also we will omit the indirect Coulomb interaction
between conduction electrons in the same lead. On one hand this contribution does not
lead to the interaction between the leads and on the other hand it leads to spin subband
splitting which is much smaller than the splitting due to the s-d interaction.

We assume that IEC is small due to large distance between the leads and
weak overlap of electron wave functions. Therefore we use the first order
perturbation theory to study this coupling. We average the Coulomb operator, $\hat H_\mathrm C$,
over the ground state of unperturbed (non-interacting) system. We use the
local coordinate spin system with z-axis being co-directed with lead magnetic moment.
Therefore, the operator of Coulomb interaction is different for FM and AFM
orientations of leads magnetic moments. The reduced Hamiltonian takes
into account only interaction between electrons at different leads and
has the form 
\begin{equation}\label{Eq:ExHam1}
\hat H_\mathrm{C}^\mathrm{ex}=
\left\{\begin{array}{l}-\sum_{i,j,s} U^{\mathrm{ss}}_{iijj}\hat a^{s+}_i\hat a^{s}_i\hat b^{s+}_j\hat b^{s}_j,~\mathrm{FM~ordering},\\ -\sum_{i,j,s} \tilde U^{\mathrm{ss}}_{iijj}\hat a^{s+}_i\hat a^{s}_i\hat b^{-s+}_j\hat b^{-s}_j,~\mathrm{AFM~ordering},\end{array}\right.\
\end{equation}
The operator does not include the factor $1/2$
because we sum over the equal elements ($U^{\mathrm{s1s2}}_{ijkl}\hat a^{s1+}_i\hat a^{s2}_j\hat b^{s2+}_k\hat b^{s1}_l$) and ($U^{\mathrm{s1s2}}_{jilk}\hat a^{s2+}_k\hat a^{s1}_l\hat b^{s1+}_i\hat b^{s2}_j$).

The electron wave function has a random phase due to scattering on impurities therefore only
the diagonal matrix elements of exchange interaction contribute
to the system energy, $U^{\mathrm{s1s2}}_{iikk}\ne 0$. Below we will omit
the double subscripts and superscripts for matrix elements of the Coulomb interaction.
The matrix element has the form~\cite{Landau3}
\begin{equation}\label{Eq:ExMatEl}
\eqalign{
&U^{s}_{ij}=\frac{1}{S}\!\int\!\!\int \rmd^3\bi r_1 \rmd^3 \bi r_2 \psi_i^{s*}(\rv_1)\phi^s_j(\rv_1)\hat U_\mathrm C \psi^s_i(\rv_2)\phi^{s*}_j(\rv_2),\cr
&\tilde U^{s}_{ij}=\frac{1}{S}\!\int\!\!\int \rmd^3\bi r_1 \rmd^3 \bi r_2 \psi_i^{s*}(\rv_1)\phi^{-s}_j(\rv_1)\hat U_\mathrm C \psi^{s}_i(\rv_2)\phi^{-s*}_j(\rv_2),
}
\end{equation}
where $\hat U_\mathrm C$ is the operator of the Coulomb interaction.
For homogeneous insulator it has the form $\hat U_{\mathrm{C}}=e^2/(4\pi\varepsilon_0\varepsilon |\rv_1-\rv_2|)$, where $\varepsilon$
is the medium effective dielectric constant. In our case the system is
inhomogeneous and the Coulomb interaction is renormalized due to screening
effects due to presence of metallic leads.

The right hand side of  ~(\ref{Eq:ExMatEl}) can be considered as the Coulomb interaction between two effective charges, $\rho_{ij}^{(1)}=e\psi^{s*}_i(\rv)\phi^{s'}_j(\rv)$ and $\rho_{ij}^{(2)}=e\psi^s_i(\rv)\phi^{s'*}_j(\rv)$. Here we use $s'=s$ for FM and $s'=-s$ for AFM configurations.

The wave functions of two electrons located at different
leads $\psi_i$ and $\phi_j$ are overlapped inside the insulating layer and
inside the leads. Therefore, there are two regions contributing to  ~(\ref{Eq:ExMatEl}):

1) The region inside the FM leads $\Omega_{1}$ ($\Omega_2$) where the Coulomb interaction is effectively
screened and is short-range ~\cite{Glazman2002}
\begin{equation}\label{Eq:CoulLeads}
\hat U^\mathrm L_\mathrm C=\frac{\Omega\Delta}{2}\delta(\bi r_1-\bi r_2),
\end{equation}
where $\Delta$ is the mean energy level spacing, $\Omega\Delta=6\pi^2E_\mathrm F/((k_\mathrm F^+)^3+(k_\mathrm F^-)^3)$,
with the Fermi momentum $k_\mathrm F^s=\sqrt{2m_\mathrm e(U+E_\mathrm F-s\Ji)/\hbar^2}$.
The peculiarity of this term is related to the fact that it does not depend
on dielectric properties of insulating layer. This region gives the following contribution
\begin{equation}\label{Eq:ExMatEl21}
L^s_{ij}=\frac{1}{S}\!\int\!\!\int_{\Omega_1+\Omega_2} \!\!\!\!\rmd^3\bi r_1 \rmd^3 \bi r_2 \psi_i^{s*}(\rv_1)\phi^{s}_j(\rv_1)\hat U^\mathrm L_\mathrm C \psi^s_i(\rv_2)\phi^{s*}_j(\rv_2),
\end{equation}
where $\Omega_{1,2}=\Omega$ is the lead volume.

2) The second region contributing to  ~(\ref{Eq:ExMatEl}) is the region
between the leads where screening of the Coulomb interaction is
weak and the interaction is long-range.
However, due to metallic leads, the electric field of two interacting
electrons is finite only inside this region. We denote the renormalized Coulomb
interaction inside the insulating layer as $\hat U^\mathrm I_\mathrm C$.
This contribution depends on the dielectric permittivity $\varepsilon$ of
the insulating layer. This region gives the following contribution
\begin{equation}\label{Eq:ExMatEl22}
I^s_{ij}=\frac{1}{S}\!\int\!\!\int_{\Omega_\mathrm I} \rmd^3\bi r_1 \rmd^3 \bi r_2 \psi_i^{s*}(\rv_1)\phi^{s}_j(\rv_1)\hat U^\mathrm{I}_\mathrm C \psi^s_i(\rv_2)\phi^{s*}_j(\rv_2),
\end{equation}
where $\Omega_\mathrm I$ is the volume of the insulating layer. We can write the matrix elements of the indirect Coulomb interaction as a sum of two terms.
For FM ordering we have
\begin{equation}\label{Eq:ExMatEl2}
U^{s}_{ij}=L^s_{ij}+I^s_{ij}.
\end{equation}
Note that electrons inside the insulator and electrons inside the leads are decoupled and
do not interact with each other. Therefore, there are no terms with integration over the
$\bi{r}_1$ in the region $\Omega_\mathrm I$ and over the $\bi{r}_2$ in the
regions $\Omega_{1,2}$ ($\int_{\Omega_\mathrm I} \rmd^3\bi r_1  \int_{\Omega_{1,2}} \rmd^3 \bi r_2$).

Similarly we can write the matrix elements for AFM state $\tilde U^s_{ij}$.
We can split the total IEC based on the Coulomb interaction into two contributions
\begin{equation}\label{Eq:Ex2}
H^\mathrm{ex}_{\mathrm C}=L^\mathrm{ex}+I^\mathrm{ex}.\\
\end{equation}
Below we consider these two contributions to the Coulomb based exchange interaction separately.

\subsection{Contribution to the exchange interaction due to the insulating region, $I^\mathrm{ex}$}

This contribution includes three terms:
1) $\tilde I^-_\mathrm{ex}$ -  FM configuration, majority spin subband;
2) $\tilde I^+_\mathrm{ex}$ -  FM configuration, minority spin subband;
3) $\tilde{\tilde I}_\mathrm{ex}$ - AFM configuration (both spin subbands give
the same contribution in AFM case).

Consider the first term  $\tilde I^-_\mathrm{ex}$.
Replacing summation in  ~(\ref{Eq:ExHam1}) with integration we have
\begin{equation}\label{Eq:ExEnEst2}
\tilde I^-_\mathrm{ex}=-\frac{\Omega^{2}}{(2\pi)^6S}\int\int_{\epsilon^-_1,\epsilon^-_2<E_\mathrm F} \!\! \rmd^3k_{1} \rmd^3k_{2} I^{\mathrm{-}}_{\bi k_1 \bi k_2},
\end{equation}
where $\epsilon^s=\hbar^2k^2/(2m_\mathrm e)+s\Ji+U$.
Substituting  ~(\ref{Eq:WaveFunc2}) and  ~(\ref{Eq:WaveFunc3}) into  ~(\ref{Eq:ExEnEst2}) we obtain
\begin{equation}\label{Eq:ExEnEst3}
\eqalign{
\tilde I^-_\mathrm{ex}= \frac{-\Omega^{2}}{(2\pi)^6S}\int \rmd^3 r_1 \rmd^3 r_2 \rmd^3k_{1} \rmd^3k_{2}\hat U^\mathrm I_\mathrm C(\bi r_1,\bi r_2) \times\cr \times \rho(z_1)(\rho(z_2))^*\rme^{\rmi(\bi k_{1\perp}-\bi k_{2\perp})\bi r_{1\perp}}\rme^{-\rmi(\bi k_{1\perp}-\bi k_{2\perp})\bi r_{2\perp}},
}
\end{equation}
where
\begin{equation}\label{Eq:ExEnEst4}
\rho(z)=\frac{e\tau_{i}^{s*}\tau^{s}_{j}}{\Omega}\rme^{ -z(\varkappa^s_{i}-\varkappa^{s}_{j})}\rme^{-\frac{d}{2}(\varkappa^s_i+\varkappa^{s}_j)}.
\end{equation}

The integral over $\bi r_{1,2}$ describes the interaction energy of
two effective charges. The effective charges decay
exponentially along z direction and harmonically vary in the transverse
direction. The effective charges produce an electric field. This field is screened
by metallic leads and is finite only inside the insulating layer. The
details of calculations are shown in \ref{App:Calc}. The final result
for $\tilde I^-_\mathrm{ex}$ has the form
\begin{equation}\label{Eq:ExEnEst8}
\eqalign{
\tilde I^-_\mathrm{ex}= -\frac{e^2}{32\pi^4\varepsilon_0\varepsilon} \int_{0}^{k^-_\mathrm F}\int_{0}^{k_1}\!\!\! \rmd k_{1}\rmd k_{2} |(\tau^-_1)^*\tau^-_2|^2 \rme^{-d(\varkappa^-_1+\varkappa^-_2)} \times \cr \times \int_{0}^{k_2^{\mathrm{max}}+k_1^{\mathrm{max}}}q \omega_\mathrm I (q) \rmd q \!\!\int_0^{(k_2^{\mathrm{max}}+k_1^{\mathrm{max}})/2}\!\!\! k\zeta(k,q)\rmd k,
}
\end{equation}
where $\varkappa$ in the expression for $\omega_\mathrm I(q)$ (\ref{Eq:InPlaneInt6}) is taken for
majority ($s=$``-'') spin subband and we introduce the notation
$k_1^{\mathrm{max}}=\sqrt{(k^-_\mathrm F)^2 - k^2_{1z}}$ and $k_2^{\mathrm{max}}=\sqrt{(k^-_\mathrm F)^2 - k^2_{2z}}$.
The functions $\omega_\mathrm I(q)$ and $\zeta(k,q)$ are defined in   \ref{App:Calc}.

The contribution from the minority ($s=$``+'') spin subband has the form
\begin{equation}\label{Eq:ExEnEst81}
\eqalign{
\tilde I^+_\mathrm{ex}= -\frac{e^2}{32\pi^4\varepsilon_0\varepsilon} \int_{0}^{k^+_\mathrm F}\int_{0}^{k_1}\!\!\! \rmd k_{1}\rmd k_{2} |(\tau^+_1)^*\tau^+_2|^2 \rme^{-d(\varkappa^+_1+\varkappa^+_2)} \times \cr \times \int_{0}^{k_2^{\mathrm{max}}+k_1^{\mathrm{max}}}q \omega_\mathrm I (q) \rmd q \!\!\int_0^{(k_2^{\mathrm{max}}+k_1^{\mathrm{max}})/2}\!\!\! k\zeta(k,q)\rmd k.
}
\end{equation}
We use $\varkappa$ for minority spin subband in the expression for $\omega_\mathrm I(q)$
and $k^\mathrm{max}_{1,2}$ should be taken for minority spin subband ($k^-_\mathrm F\to k^+_\mathrm F$).

Similarly we calculate the contribution due to AFM configuration
\begin{equation}\label{Eq:ExEnEst9}
\eqalign{
 \tilde{\tilde I}_\mathrm{ex}=-\frac{e^2}{32\pi^4\varepsilon_0\varepsilon} \int_{0}^{k^+_\mathrm F}\int_{0}^{k^-_\mathrm F}\!\!\! \rmd k_{1}\rmd k_{2} |(\tau^+_1)^*\tau^-_2|^2 \rme^{-d(\varkappa^+_1+\varkappa^-_2)} \times \cr \times \int_{0}^{k_2^{\mathrm{max}}+k_1^{\mathrm{max}}}q \omega_\mathrm I (q) \rmd q \!\!\int_0^{(k_2^{\mathrm{max}}+k_1^{\mathrm{max}})/2}\!\!\! k\zeta(k,q)\rmd k.
}
\end{equation}
Here we use $\varkappa_2$ for majority spin subband and $\varkappa_1$ for minority
spin subband in the expression for $\omega_\mathrm I$. Similarly, we
have $k_1^{\mathrm{max}}=\sqrt{(k^+_\mathrm F)^2 - k^2_{1z}}$ and $k_2^{\mathrm{max}}=\sqrt{(k^-_\mathrm F)^2 - k^2_{2z}}$.
The contributions to the energy of AFM configuration from both spin
subbands are equal and were included into $\tilde{\tilde I}_\mathrm{ex}$.
Using  ~(\ref{Eq:ExEnEst8}-\ref{Eq:ExEnEst9}) we find the total
contribution to the exchange interaction due to the insulating region
\begin{equation}\label{Eq:ExInsTot}
I^\mathrm{ex}=\tilde{\tilde I}_\mathrm{ex}-\tilde I_\mathrm{ex}^+-\tilde I_\mathrm{ex}^-.
\end{equation}

\subsection{Contribution to the exchange interaction due to the FM leads, $L^\mathrm{ex}$}

Inside the metallic leads the Coulomb interaction is short-range therefore the matrix element of the
exchange interaction has the form
\begin{equation}\label{Eq:ExLEads}
L^s_{ij}=\frac{\Omega\Delta}{2S}\!\int_{\Omega_1+\Omega_2} \!\!\rmd^3\bi r |\psi_i^{s}(\rv)|^2|\phi^{s'}_j(\rv)|^2,
\end{equation}
where $s'=s$ for FM and $s'=-s$ for AFM states.
Using  ~(\ref{Eq:WaveFunc1}) and~(\ref{Eq:WaveFunc2}) we obtain for normalized matrix element the following result
\begin{equation}\label{Eq:ExLEads1}
\eqalign{
L^s_{ij}=\frac{\Delta  |\tau^s_{j}|^2\rme^{-2d\varkappa^s_{j}}}{\Omega}\left(\frac{1+|\xi^{s'}_{i}|^2}{2\varkappa_j^s}+\mathrm{Re}\left(\frac{(\xi^{s'}_i)^*}{\varkappa^s_j+\rmi k_i}\right)\right)+\cr
\frac{\Delta  |\tau^{s'}_{i}|^2\rme^{-2d\varkappa^{s'}_{i}}}{\Omega}\left(\frac{1+|\xi^{s}_{j}|^2}{2\varkappa_i^{s'}}+\mathrm{Re}\left(\frac{(\xi^{s}_j)^*}{\varkappa^{s'}_i+\rmi k_j}\right)\right).
}
\end{equation}
Integrating this matrix element over all eigenstates of the ground state we find for FM configuration
\begin{equation}\label{Eq:ExLEadsFM}
\eqalign{
\tilde L_\mathrm{ex}^s=\frac{3(U+E_\mathrm F)}{2^6\pi((k_\mathrm F^+)^3+(k_\mathrm F^-)^3))}\times\cr \times\int_0^{k_\mathrm F^s}\rmd k_1\int_0^{k_\mathrm{F}^s} \rmd k_2((k_\mathrm F^s)^2-k_2^2)((k_\mathrm F^s)^2-k_1^2)\times\cr \times\left\{\rme^{-2d\varkappa^s_{1}}|\tau^s_{1}|^2\left(\frac{1+|\xi^{s}_{2}|^2}{2\varkappa_1^s}+\mathrm{Re}\left(\frac{(\xi^{s}_{2})^*}{\varkappa^s_{1}+\rmi k_2}\right)\right)+\right.\cr
\left. |\tau^{s}_{2}|^2\rme^{-2d\varkappa^{s}_{2}}\left(\frac{1+|\xi^{s}_{1}|^2}{2\varkappa_2^s}+\mathrm{Re}\left(\frac{(\xi^{s}_1)^*}{\varkappa^{s}_2+\rmi k_1}\right)\right)\right\},
}
\end{equation}
and for AFM configuration we have
\begin{equation}\label{Eq:ExLEadsAFM}
\eqalign{
\tilde{\tilde L}_\mathrm{ex}=\frac{6(U+E_\mathrm F)}{2^6\pi((k_\mathrm F^+)^3+(k_\mathrm F^-)^3))}\times\cr \times\int_0^{k_\mathrm F^+}dk_1\int_0^{k_\mathrm{F}^-} \rmd k_2((k_\mathrm F^-)^2-k_2^2)((k_\mathrm F^+)^2-k_1^2)\times\cr \times\left\{\rme^{-2d\varkappa^+_{1}}|\tau^+_{1}|^2\left(\frac{1+|\xi^{-}_{2}|^2}{2\varkappa_1^+}+\mathrm{Re}\left(\frac{(\xi^{-}_{2})^*}{\varkappa^+_{1}+\rmi k_2}\right)\right)+\right.\cr
\left. |\tau^{-}_{2}|^2\rme^{-2d\varkappa^{-}_{2}}\left(\frac{1+|\xi^{+}_{1}|^2}{2\varkappa_2^-}+\mathrm{Re}\left(\frac{(\xi^{+}_1)^*}{\varkappa^{-}_2+\rmi k_1}\right)\right)\right\}.
}
\end{equation}
Using  ~(\ref{Eq:ExLEadsFM}) and (\ref{Eq:ExLEadsAFM}) we find the total
contribution to the exchange interaction due to the leads
\begin{equation}\label{Eq:ExLEadsTot}
L^\mathrm{ex}=\tilde{\tilde L}_\mathrm{ex}-\tilde L_\mathrm{ex}^+-\tilde L_\mathrm{ex}^-.
\end{equation}

\subsection{Total exchange interaction}

The total exchange interaction is given by the expression
\begin{equation}\label{Eq:ExTot}
H^\mathrm{ex}=H^\mathrm{ex}_\mathrm{h}+\tilde{\tilde I}_\mathrm{ex}-\tilde I^+_\mathrm{ex}-\tilde I^-_\mathrm{ex}+\tilde{\tilde L}_\mathrm{ex}-\tilde L^+_\mathrm{ex}-\tilde L^-_\mathrm{ex}.
\end{equation}
The first term was calculated in~\cite{Bruno1995}
\begin{equation}\label{Eq:ExSl}
\eqalign{
H^\mathrm{ex}_\mathrm{h}=-\frac{\hbar^2}{\pi^2 m_\mathrm e d^2} \, \varkappa_0^5b \, \rme^{-2\varkappa_0 d},\cr
b=\left\{\begin{array}{l}\frac{(\varkappa_0^2-k_\mathrm F^+ k_\mathrm F^-)(k_\mathrm F^+ - k_\mathrm F^-)^2(k_\mathrm F^+ + k_\mathrm F^-)}{(\varkappa^2_0+(k_\mathrm F^+)^2)^2(\varkappa^2_0+(k_\mathrm F^-)^2)^2},~U+E_\mathrm F>\Ji,\cr
\frac{k_\mathrm F^-((k_\mathrm F^-)^2+\rmi \varkappa_0 k_\mathrm F^+)}{\varkappa_0(\varkappa^2_0+(k_\mathrm F^-)^2)^2(\varkappa_0-\rmi k_\mathrm F^+)},~U+E_\mathrm F<\Ji.\end{array}\right.
}
\end{equation}
Here the coefficient $b$ is always real. For $U+E_\mathrm F<\Ji$ (the second line)
the Fermi momentum $k_\mathrm F^+$ is imaginary, $k_\mathrm F^+=\rmi |k^+_\mathrm F|$.

\subsection{Half metals}

For half metals with one spin subband ($U+E_\mathrm F<\Ji$) the first
contribution in  ~(\ref{Eq:ExGen}) is absent and only the majority
spin subband contributes to the second term in  ~(\ref{Eq:ExGen}).

\subsection{Analytical expression for limiting case}

For weak spin subband splitting, $J_\mathrm{sd}\ll (U+E_\mathrm F)$ and
large decay length, $\varkappa_0\ll k_\mathrm F$, the IEC is
quadratic in $J_\mathrm{sd}$ and can be written in powers of small
parameter $\varkappa_0/k_\mathrm F$. In the leading order we find
\begin{equation}\label{Eq:ExHopApp}
H^\mathrm{ex}_\mathrm{h}\approx -\frac{2\hbar^2}{\pi^2 m_\mathrm e d^2}\frac{J^2_\mathrm{sd}}{(U+E_\mathrm F)^2}\frac{\varkappa_0^5}{k^3_\mathrm F}\rme^{-2\varkappa_0 d},
\end{equation}
and for the Coulomb based IEC we have
\begin{equation}\label{Eq:ExCoulApp1}
H^\mathrm{ex}_\mathrm{C}\approx \frac{\zeta e^2}{4\pi\varepsilon_0\varepsilon d}\frac{J^2_\mathrm{sd}}{(U+E_\mathrm F)^2}\frac{k_\mathrm F}{d}\rme^{-2\varkappa_0 d},
\end{equation}
with $\zeta\approx 1/150$. The expression for the Coulomb based IEC is derived by fitting  ~(\ref{Eq:ExGen}).
In this limit the Coulomb based contribution gives the largest contribution to the IEC.

\section{Discussion of results}\label{Sec:Discuss}

\subsection{Comparison of three contributions to the total exchange interaction}

We split the total exchange interaction into three components $H^\mathrm{ex}_\mathrm{h}$, $L^\mathrm{ex}$ and $I^\mathrm{ex}$.
These contributions have different origins and behavior.
The first contribution, $H^\mathrm{ex}_\mathrm{h}$, appears due to spin current
between the leads. The second, $L^\mathrm{ex}$, and the third, $I^\mathrm{ex}$, contributions are
due to many-body effects. It is important that $I^\mathrm{ex}$ depends on the dielectric permittivity
of the insulating layer in contrast to the first and the second contributions.
\begin{figure}
\includegraphics[width=0.6\columnwidth]{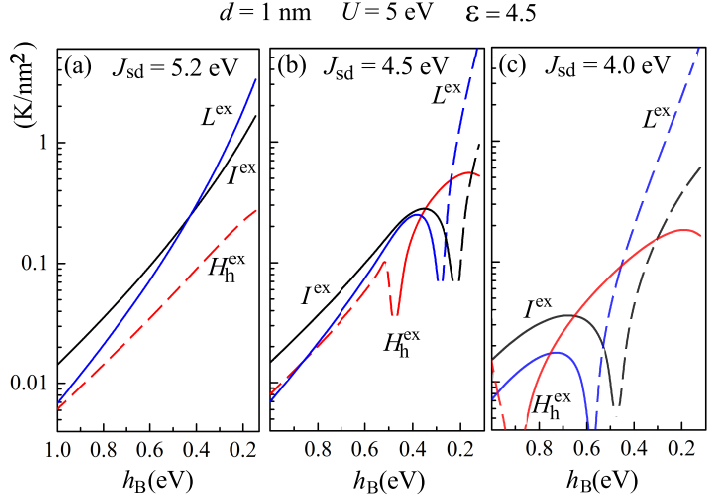}
\caption{(Color online) The interlayer exchange interaction as a function
of insulating barrier height $h_\mathrm B$ for $U=5$ eV, $\varepsilon=4.5$, $d=1$ nm and (a) $\Ji=5.2$ eV, (b) $\Ji=4.5$ eV, (c) $\Ji=4$ eV.
Black lines show $|I^\mathrm{ex}|$ (\ref{Eq:ExInsTot}),
blue lines are for $|L^\mathrm{ex}|$ (\ref{Eq:ExLEadsTot})
and red lines are for $|H^\mathrm{ex}_\mathrm{h}|$ (\ref{Eq:ExSl}). The y-axis has logarithmic scale.
Dashed parts show the region where functions
$I^\mathrm{ex}$, $L^\mathrm{ex}$ and $H^\mathrm{ex}_\mathrm{h}$ are negative.} \label{Fig:JvsEf1}
\end{figure}

Figure~\ref{Fig:JvsEf1} shows the dependence of three contributions on the
barrier height $h_\mathrm B$, in our notations $h_\mathrm B=-E_\mathrm F$, for the following parameters:
$U=5$ eV, $d=1$ nm, $\varepsilon=4.5$, $\Ji=5.2;4.5;4.0$ eV.
All three contributions have different behavior as a function of barrier height $h_B$.
For large spin subband splitting, $\Ji$ (half metals) shown in figure~\ref{Fig:JvsEf1}(a)
all three contributions do not change their sing as a function of $h_\mathrm B$. The Coulomb based IEC is positive while the hopping based one is negative. For small $\Ji$ (two spin subband metals) all three
contributions are non-monotonic and change their sign and can be either of FM or AFM type
depending on the barrier height. The contribution $I^\mathrm{ex}$ (black line)
depends on the dielectric permittivity of the insulating layer in contrast to two other contributions.
We use SiO$_2$ with $\varepsilon=4.5$ as an example. For insulators with high dielectric
constants $\varepsilon>100$ this contribution is suppressed.
\begin{figure}
\includegraphics[width=0.6\columnwidth]{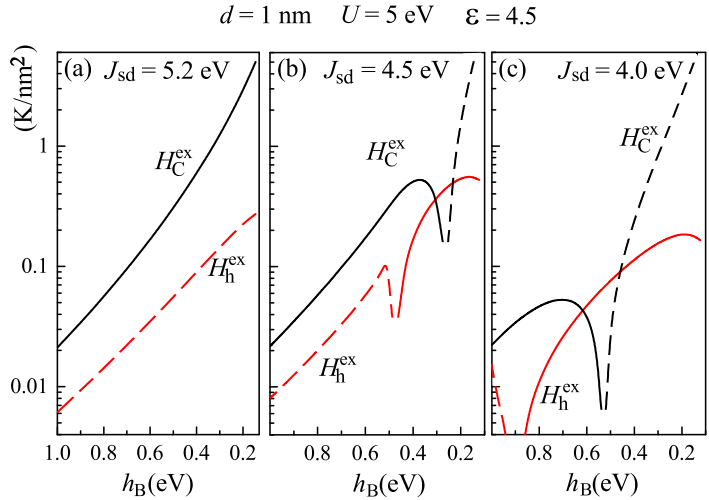}
\caption{(Color online) The interlayer exchange interaction as a function of $h_\mathrm B$ for $U=5$ eV, $\varepsilon=4.5$, $d=1$ nm and (a) $\Ji=5.2$ eV, (b) $\Ji=4.5$ eV, (c) $\Ji=4$ eV. Black lines are for $|H^\mathrm{ex}_\mathrm{C}|$ (\ref{Eq:Ex2}),
and red lines are for $|H^\mathrm{ex}_\mathrm{h}|$ (\ref{Eq:ExSl}).
Dashed parts show the region where functions $H^\mathrm{ex}_\mathrm{C}$ and $H^\mathrm{ex}_\mathrm{h}$ are negative. } \label{Fig:JvsEf2}
\end{figure}

Figure~\ref{Fig:JvsEf1} shows that all three contributions to the IEC are comparable and
there is no a single component dominating in the whole range
of parameters, especially if we take into account the fact that all three components change
their sign at different values of barrier height.  For strong subband splitting (half metal)
and $\varepsilon<4$ the Coulomb based contribution dominates for any barrier heights.
This contribution also dominates for small spin subband splitting in the region
of high barrier height ($h_\mathrm B>0.3$ eV for $\Ji=4.5$ eV and
$h_\mathrm B>0.6$ eV for $\Ji=4.0$ eV). For spin subband splitting smaller than the average
conduction band width, $W_\mathrm{con}=U+E_\mathrm F$, and for low barrier height
($h_\mathrm B<0.3$ eV in figure~\ref{Fig:JvsEf1}) the Coulomb based exchange interaction, $L^\mathrm{ex}$
provides the main contribution to the IEC. For spin subband splitting smaller
than the average conduction band width, $W_\mathrm{con}$ there is also a region where
the hopping based exchange interaction,  $H^\mathrm{ex}_\mathrm{h}$ dominates.
Using the fact that $I^\mathrm{ex}$ depends on the dielectric constant, in experiment
one can have any relation between these three contributions.

Figure~\ref{Fig:JvsEf2} shows the dependence of the total Coulomb based exchange contribution,
$H^\mathrm{ex}_\mathrm{C}=I^\mathrm{ex}+L^\mathrm{ex}$ and the hopping based contribution,
$H^\mathrm{ex}_\mathrm{h}$ as a function of barrier height, $h_\mathrm B$.
For half-metal (panel (a)) the contributions $H^\mathrm{ex}_\mathrm{h}$ and $H^\mathrm{ex}_\mathrm{C}$ have
the opposite sign for any $h_\mathrm B$. Also for full 
metal ($J_\mathrm{sd}<U+E_\mathrm F$) the hopping and the Coulomb based contributions have 
the opposite sign. However, the FM/AFM transitions appear at different barrier hight, $h_\mathrm B$. 
The FM/AFM transition point of the Coulomb based
contribution is located above the transition point of the hopping based exchange.
The Coulomb based exchange mostly exceeds the hoping based exchange.
Only in the vicinity of the FM/AFM transition point of $H^\mathrm{ex}_\mathrm{C}$ the hopping based exchange
dominates.
\begin{figure}
\includegraphics[width=0.6\columnwidth]{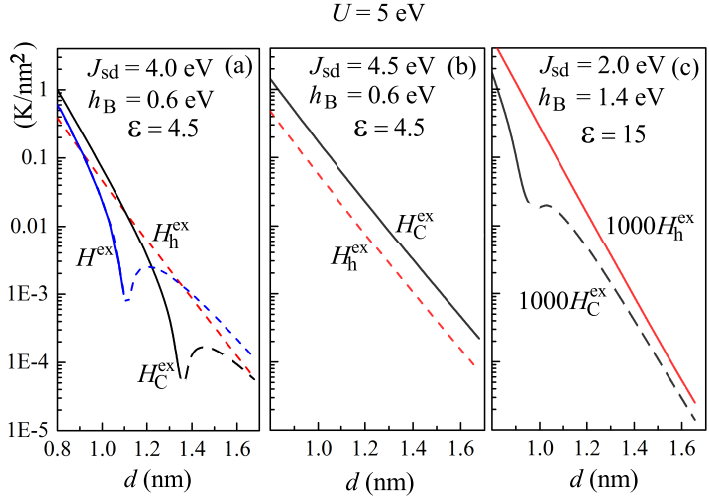}
\caption{(Color online) The interlayer exchange interaction as a function of the insulator thickness $d$ for $U=5$ eV, and (a) $\Ji=4.0$ eV, $h_\mathrm B=0.6$ eV, $\varepsilon=4.5$, (b) $\Ji=4.5$ eV, $h_\mathrm B=0.6$ eV, $\varepsilon=4.5$, (c) $\Ji=2.0$ eV, $h_\mathrm B=1.4$ eV, $\varepsilon=15$. Black lines are for $|H^\mathrm{ex}_\mathrm{C}|$ (\ref{Eq:Ex2}), red lines are for $|H^\mathrm{ex}_\mathrm{h}|$ (\ref{Eq:ExSl}),
blue line is for total exchange interaction, $H^\mathrm{ex}$ (\ref{Eq:ExTot}).
Dashed parts show the region where functions are negative.} \label{Fig:Jvsd}
\end{figure}

Figure~\ref{Fig:Jvsd} shows the behavior of the IEC as a function of the
insulator thickness $d$ for the following parameters $U=5$ eV, $\varepsilon=4.5;15$.
The Coulomb and the hopping based contributions
are shown separately. The total exchange interaction is also shown.
The barrier height and the spin subband splitting is different for different panels.
Panel (b) ($\Ji=4.5$ eV and $h_\mathrm B=0.6$ eV, half-metal leads) shows
the case where the Coulomb exchange dominates. In this case the exchange coupling
decays exponentially with thickness $d$ and has the same decay rate
as the hopping based exchange ($\rme^{-2d\varkappa_0}/d^2$).
Therefore the dependence on thickness does not allow to distinguish the Coulomb and
the hopping based exchange contributions.

Panel (c) ($\Ji=2.0$ eV and $h_\mathrm B=1.4$ eV, two spin subband metal) shows
the opposite situation when the hoping based exchange coupling dominates.
Figure~\ref{Fig:Jvsd}(a) ($\Ji=4$ eV, $h_\mathrm B=0.6$ eV) shows the case
when both contributions are of the same order. In this situation the Coulomb
based exchange interaction has a more complicated behavior which is
substantially different from the behavior of the hopping based exchange contribution.
At some thickness
the Coulomb based contribution changes its sign. For small thickness the Coulomb
based contribution $H^\mathrm{ex}_\mathrm C$ is positive (FM) while the hopping
based exchange is negative (AFM). For large thickness both contributions are
of AFM type. The blue curve shows the total IEC. One can see that $H^\mathrm{ex}$
also changes its sign. Such a behavior is due to competition of
two contributions. The Coulomb based exchange interaction can be studied using
this non-monotonic behavior.
\begin{figure}
\includegraphics[width=0.6\columnwidth]{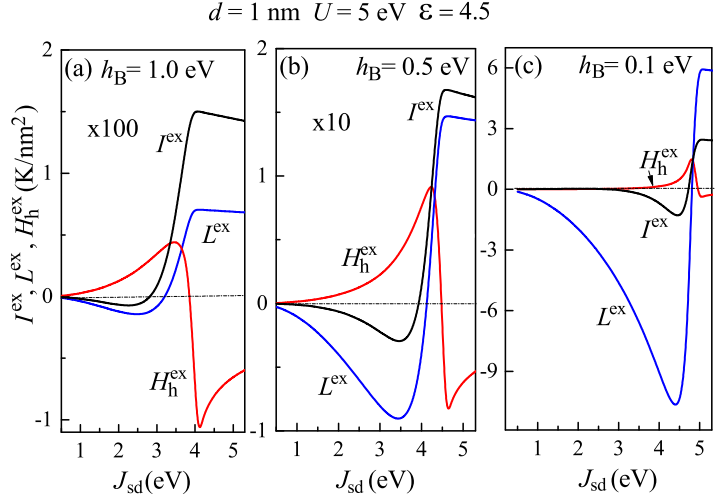}
\caption{(Color online) The interlayer exchange interaction as a function of
spin subband splitting, $\Ji$, for $U=5$ eV, $\varepsilon=4.5$, $d=1$ nm,
and (a) $h_\mathrm B=1.0$ eV, (b) $h_\mathrm B=0.5$ eV, (c) $h_\mathrm B=0.1$ eV.
Black lines are for $I^\mathrm{ex}$ (\ref{Eq:ExInsTot}),
blue line is for $L^\mathrm{ex}$ (\ref{Eq:ExLEadsTot}), and
red lines are for $H^\mathrm{ex}_\mathrm{h}$ (\ref{Eq:ExSl}).
The data are multiplied by 100 in
panel (a) and by 10 in panel (b).} \label{Fig:JvsJ1}
\end{figure}

Figure~\ref{Fig:JvsJ1} shows the dependence of the exchange coupling on the spin subband splitting $\Ji$ for $U=5$ eV, $\varepsilon=4.5$, $d=1$ nm, and $h_\mathrm B=1.0;0.5;0.1$ eV. These plots show that contribution $I^\mathrm{ex}$ dominates
if only one spin subband is filled ($\Ji>U+E_\mathrm F$).
For small splitting
the main contribution is due to hopping, $H^\mathrm{ex}_\mathrm{h}$ and the Coulomb based, $L^\mathrm{ex}$ exchange.
Figure~\ref{Fig:JvsJ1} shows that transition from FM to AFM coupling for different contributions
appears at different values of the spin subband splitting, $\Ji$. In addition, the behavior
of all contributions strongly depends on the barrier height.

\subsection{Total IEC vs spin subband splitting $\Ji$, barrier height $h_\mathrm B$ and insulator thickness $d$}

\begin{figure}
\includegraphics[width=0.6\columnwidth]{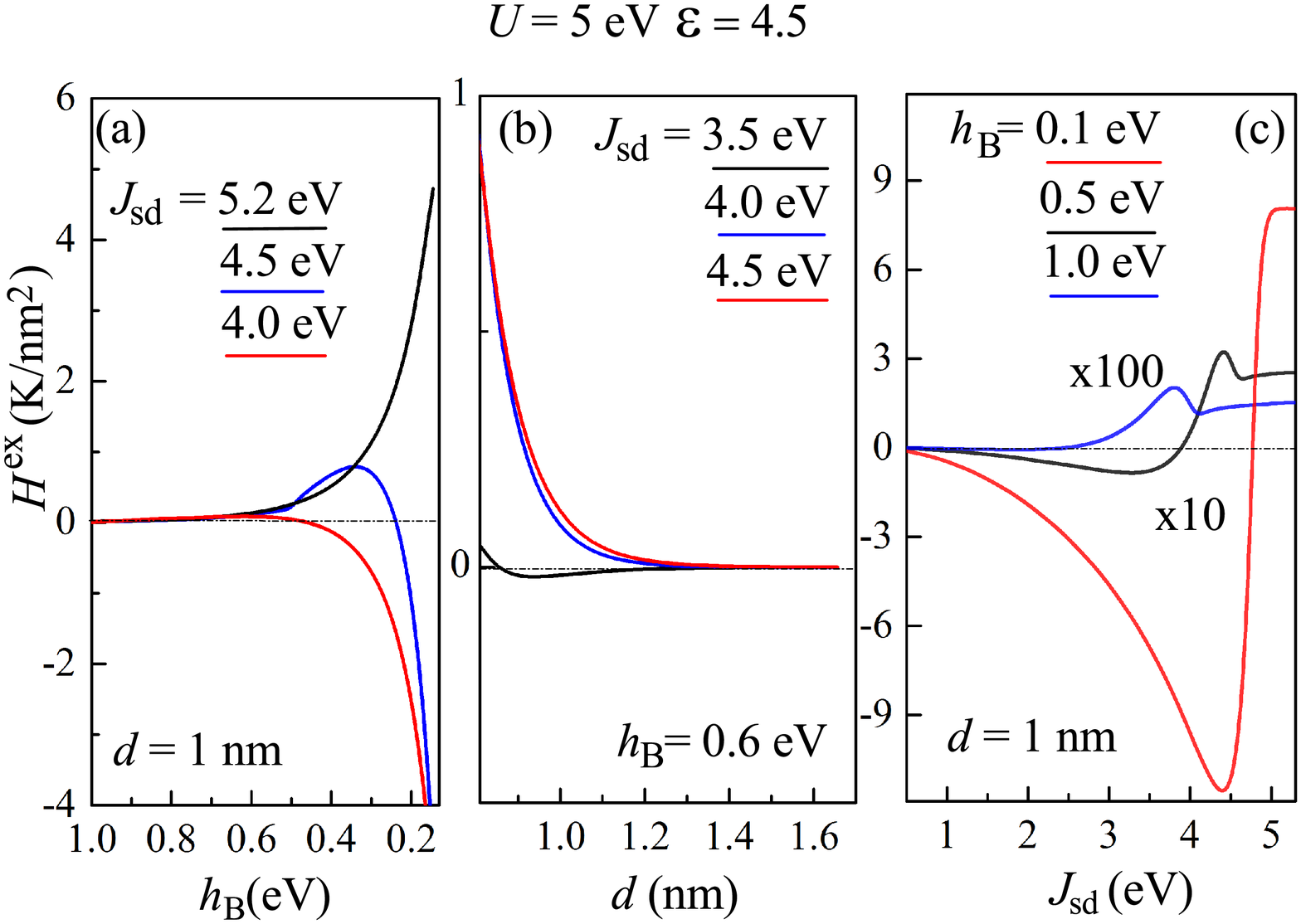}
\caption{(Color online) Total interlayer exchange coupling (IEC), $H^\mathrm{ex}$ in
 (\ref{Eq:ExTot}) as a function of (a) barrier height $h_\mathrm B$;
(b) barrier thickness $d$; (c) and the spin subband splitting $\Ji$ for $U=5$ eV, $\varepsilon=4.5$.
In panel (c) the data for $h_\mathrm B=1.0$ eV are multiplied by 100 and for $h_\mathrm B=0.5$ eV  by 10.} \label{Fig:ExTot}
\end{figure}

Figure~\ref{Fig:ExTot} shows the total IEC,  $H^\mathrm{ex}$ in  ~(\ref{Eq:ExTot})
as a function of barrier height $h_\mathrm B$, barrier thickness $d$, and the spin subband splitting $\Ji$
for $U=5$ eV, $\varepsilon=4.5$.  These curves show that $H^\mathrm{ex}$ 
 decays with the thickness of the insulating layer
and decays with increasing the barrier height. For small spin subband splitting the coupling is weak
and negative (AFM). For large subband splitting the IEC becomes FM (positive)
and reaches its maximum value.  The important feature of the Coulomb based exchange
coupling is its dependence on the dielectric constant $\varepsilon$ of insulator layer.
The hopping based exchange coupling does not depend on $\varepsilon$.

\subsection{Coulomb based exchange coupling vs dielectric permittivity of the insulating layer: ME effect}

\begin{figure}
\includegraphics[width=0.6\columnwidth]{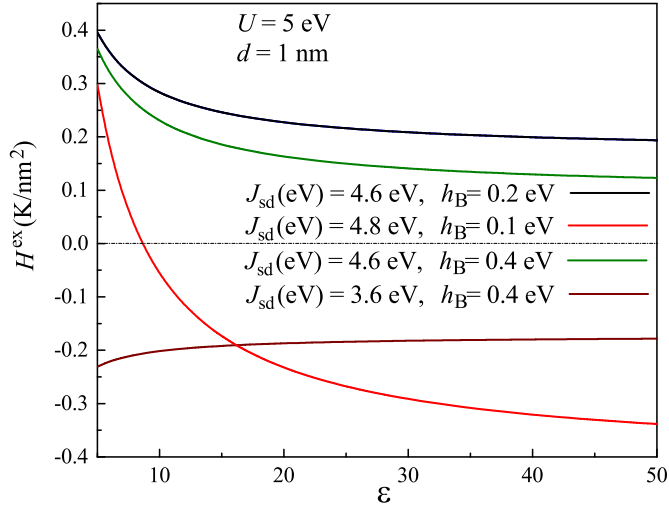}
\caption{(Color online) Total interlayer exchange coupling
(IEC), $H^\mathrm{ex}$ in (\ref{Eq:ExTot}) as a function of dielectric
permittivity of the insulating layer, $\varepsilon$, for $U=5$ eV, $d=1$ nm and
different spin subband spitting, $\Ji$, and barrier height $h_\mathrm B$.} \label{Fig:JvsEps}
\end{figure}

The important feature of the Coulomb based exchange interaction
is its dependence on the dielectric permittivity $\varepsilon$ of the insulating layer.
The Coulomb based exchange interaction has two contributions:
1) The indirect Coulomb interaction inside the leads - $L^\mathrm{ex}$ in  ~(\ref{Eq:ExLEadsTot}).
This term does not depend on the properties of the insulating layer. 2) The indirect Coulomb interaction
inside the insulating layer - $I^\mathrm{ex}$ in  ~(\ref{Eq:ExInsTot}).
This term is inversely proportional to the dielectric constant of the insulating layer,
$I^\mathrm{ex}\sim \varepsilon^{-1}$. To observe a strong dependence
of the total IEC, $H^\mathrm{ex}$, on $\varepsilon$ the
MTJ should be made of materials with large $I^\mathrm{ex}$ contribution, larger than two other contributions.
Figures~\ref{Fig:JvsEf1} and \ref{Fig:JvsJ1} show that the Coulomb based exchange interaction
is the largest contribution for strong spin subband splitting ($\Ji>U+E_\mathrm F$) and not too small barrier height.
Figure~\ref{Fig:JvsEps} shows the total exchange as a function of the dielectric constant at different $\Ji$ and $h_\mathrm B$.
These curves can be fitted using the expression
\begin{equation}\label{Eq:ExEps}
H^\mathrm{ex}=H^\mathrm{ex}_0+\frac{I^\mathrm{ex}_1}{\varepsilon},
\end{equation}
where $H^\mathrm{ex}_0=H^\mathrm{ex}_\mathrm{h}+L^\mathrm{ex}$ is the part of the total exchange
coupling which does not depend on the dielectric permittivity and
$I^\mathrm{ex}_1$ is the Coulomb based exchange coupling inside the insulator for $\varepsilon=1$.
Depending on the barrier height $h_\mathrm B$, the
thickness $d$ of the insulator, and the spin subband splitting, the contribution $H^\mathrm{ex}_0$
can be either positive (green and black curves in figure~\ref{Fig:JvsEps}) or negative
(red and brown curves in figure~\ref{Fig:JvsEps}). For negative $H^\mathrm{ex}_0$ the total IEC 
may change its sign with increasing the dielectric permittivity of the insulating layer (see the red curve).
For small $\varepsilon$ the IEC is positive (FM) while for large $\varepsilon$ it is negative (AFM).
The dependence of $H_\mathrm C^\mathrm{ex}$ on $\varepsilon$ can be used to distinguish this contribution
from the hopping based exchange coupling. In experiment one can use an insulator with
dielectric constant depending on some external parameter. In FE the dielectric permittivity
depends on temperature and electric field. However, most FEs have rather large
dielectric constants, $\varepsilon>100$. These values of  $\varepsilon$ can suppress
the Coulomb based exchange coupling. It is important to use FE with low dielectric permittivity
such as TTF-CA~\cite{Saito1991,Tokura2012} or Hf$_{0.5}$Zr$_{0.5}$O$_2$~\cite{Mikolajick2012}.

\begin{figure}
\includegraphics[width=0.6\columnwidth]{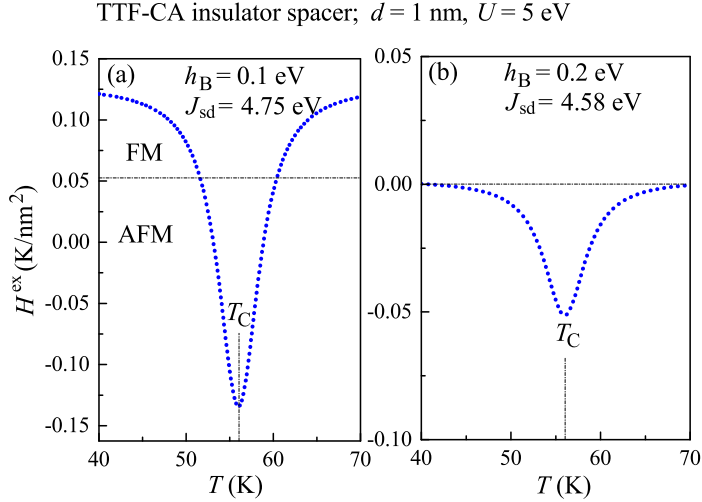}
\caption{(Color online) Total IEC, $H^\mathrm{ex}$ in  (\ref{Eq:ExTot}) as a function of temperature
for MTJ with TTF-CA FE layer ($\varepsilon$ is given by (\ref{Eq:EpsVsT})). 
The system parameters are: $U=5$ eV $d=1$ nm, and (a) $\Ji=4.75$ eV,  $h_\mathrm B=0.1$ eV;
(b) $\Ji=4.58$ eV, $h_\mathrm B=0.2$ eV. $T_\mathrm C$ is the Curie point of TTF-CA ferroelectric.} \label{Fig:JvsT1}
\end{figure}

Figure~\ref{Fig:JvsT1} shows the temperature dependence of IEC, $H^\mathrm{ex}$, with TTF-CA FE as an insulating layer.
The FE Curie point of TTF-CA is $T_\mathrm C=56$ K~\cite{Keller2014,Plank2014}.
Using data of~\cite{Keller2014,Plank2014} we find the temperature dependence of TTF-CA
dielectric constant. Experimental data are fitted well with the following dependence of dielectric constant on temperature
\begin{equation}\label{Eq:EpsVsT}
\varepsilon (T)=10+\frac{25}{1+(T-T_\mathrm C)^2/\Delta T^2},
\end{equation}
where $T_\mathrm C=56$ K is the Curie temperature of the ferroelectric-paraelectric (FE/PE)
phase transition of TTF-CA and $\Delta T=1.5$ K is the width of the peak. In the vicinity of the FE phase
transition the dielectric permittivity of TTF-CA increases leading to decrease
of the IEC. Panel (a) shows the change of sign of IEC from FM to AFM state in the vicinity
of the FE/PE phase transition.
Panel (b) shows the case of zero IEC away from the FE Curie point
and finite IEC in the vicinity of $T_\mathrm C$.

\begin{figure}
\includegraphics[width=0.6\columnwidth]{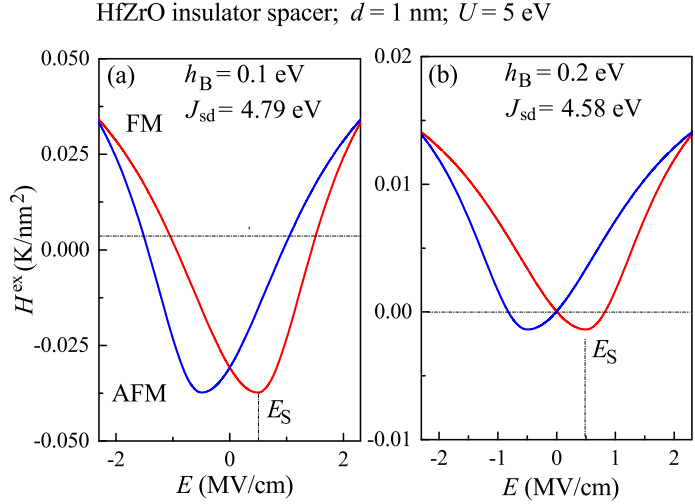}
\caption{(Color online) Total IEC $H^\mathrm{ex}$,  (\ref{Eq:ExTot}) as a function
of electric field inside of the insulating layer (voltage applied across MTJ) for MTJ
with Hf$_{0.5}$Zr$_{0.5}$O$_2$ FE layer. The system parameters
are: $U=5$ eV $d=1$ nm, and (a) $\Ji=4.79$, eV $h_\mathrm B=0.1$ eV; 
(b) $\Ji=4.58$, $h_\mathrm B=0.2$ eV. $E_\mathrm S$ is the FE polarization switching field.} \label{Fig:JvsE1}
\end{figure}

Figure~\ref{Fig:JvsE1} shows the interlayer exchange coupling, $H^\mathrm{ex}$ vs
external electric field for Hf$_{0.5}$Zr$_{0.5}$O$_2$ FE layer.
Here the voltage is applied across the MTJ with 1 nm Hf$_{0.5}$Zr$_{0.5}$O$_2$ insulating layer.
This voltage produces an electric field inside the FE. We use data of ~\cite{Mikolajick2012}
for the dependence of dielectric constant on the electric field for Hf$_{0.5}$Zr$_{0.5}$O$_2$ FE.
The curves in  figure~\ref{Fig:JvsE1} are shown at room temperature where Hf$_{0.5}$Zr$_{0.5}$O$_2$ is in the FE phase.
The dielectric permittivity $\varepsilon$ vs electric field has two branches due to hysteresis and
it has two peaks in the vicinity of the switching fields, $E=\pm E_\mathrm S$.
These two peaks correspond to two dips in the $H^\mathrm{ex}$ curves. Panel (a) shows
the change of sign of IEC with applied voltage: at zero voltage the coupling is AFM
while at larger voltages the crossover from AFM to FM coupling occurs.
Panel (b) shows the case of zero IEC at zero electric field and
finite IEC at finite electric field. Thus, this panel explicitly shows the
possible to control the IEC in MTJ using electric field.

Note that the applied voltage in FE MTJ changes not only the dielectric constant of the barrier but also 
the potential barrier itself and changes the FE polarization. These effects 
are not taken into account in our consideration but they lead to the ME effect as well~\cite{Brataas2008,Lee2009,SUZUKI2008,Ralph2008,Vedyayev2005}. The ME effect in 
the above cited work is related to the hopping based IEC effect. The IEC effect in MTJ 
without FE barrier is quadratic in voltage in symmetric 
junctions~\cite{Brataas2008,Lee2009,SUZUKI2008,Ralph2008}. This is in contrast to the 
ME effect considered in the current manuscript having linear contribution at small 
voltages. From symmetry point of view, the polarization of the FE barrier should lead to the 
occurrence of linear contribution to the ME effect related to the hopping 
based IEC in symmetric MTJ. This case, however, was not considered in the literature. 
In asymmetric MTJ (considered in~\cite{Vedyayev2005}) only the case of zero voltage 
was discussed. Due to MTJ asymmetry the polarization switching leads to variation of IEC 
even at zero voltage. A full comparison of ME effect due to the Coulomb based IEC and the
hopping based IEC is beyond the scope of the current manuscript.

The dependence of the IEC on dielectric permittivity is due to
coupling between magnetic and electric degrees of
freedom in MTJ. This magneto-electric coupling has
the Coulomb nature and allows controlling the magnetic state of MTJ.
A similar effect was predicted semi-phenomenologically
for granular magnets~\cite{Bel2014ME,Bel2014ME1,Bel2014ME2,Bel2014ME3}.

\section{Validity of our model}\label{Sec:Val}

Below we discuss several assumptions and approximations of our theory.

1) We assume that the leads are ``perfect'' metals meaning that they screen the electric field completely.
We use this assumption to calculate the electric field inside the insulating layer.
We assume that the field created by charges is zero inside the metal. Similarly we assume that
the Coulomb interaction inside the metal is a short-range and is described by the $\delta$-function.
In fact, the  electric potential created by a point charge located in a metal decays
exponentially with distance, $\sim \rme^{-r/\lambda_\mathrm{TF}}/r$, where $\lambda_\mathrm{TF}$ is
Thomas-Fermi length. The electric field created by a charge located outside the metal
also penetrates into the metal up to the distance $\lambda_\mathrm{TF}$. Therefore in the vicinity
of the metal surface (for distances less than $\lambda_\mathrm{TF}$) our consideration is not valid.
However, the Thomas-Fermi length is of the order of 0.05 nm and is much smaller than the
characteristic decay length $\varkappa_0$ of electron wave function and the thickness $d$
of the insulating layer.

Also the metal-insulator barrier is not a step function as we assumed.
The smearing of the boundary is comparable with $\lambda_\mathrm{TF}$.
In this region the classical description of screening inside the insulating barrier is not valid.
We use the method of image charges to calculate screening. This approach works well if
the distance from the metal surface is larger than the size of the exchange-correlation hole
which is less than 0.1 nm. For distances less than 0.1 nm to the metal surface a quantum theory
is required to describe the screening effects. Thus, our theory is not valid for
MTJ with mono-layer insulating barrier~\cite{Korecki2014}.

2) In section~\ref{Sec:Discuss} we discuss the ME coupling in MTJ
due to the Coulomb based exchange interaction. To observe this coupling the
FE layer should be present between magnetic leads. For strong exchange coupling the thickness of
the insulating barrier should be about 1 nm. It is known that the FE Curie temperature
decreases with decreasing the FE thickness~\cite{Frid2006rev,Frid2010rev}.
Some FEs show the critical thickness where FE properties can disappear.
However, there are FEs that can keep their properties down to a single atomic layer
thickness~\cite{Frid2006rev,Frid2010rev}. In this paper we have not discussed
the influence of size effects on the FE properties of insulating layer.

3) The voltage across the MTJ, necessary for creation of electric field in the insulating barrier,
causes the electric and spin currents through the barrier. Such a current produces a spin transfer
torque leading to the magnetic interaction between FM leads. This effect was not discussed here.

4)\textit{Ab initio} calculations based on density functional theory mostly use
the local spin density (LSD) approximation treating the indirect Coulomb interaction (exchange)
as the short-range interaction, $\hat H_\mathrm C\sim \delta(\bi r_1-\bi r_2)$.
This approach is valid only if the characteristic scales of the electron
density variations are much larger than the screening radius and the
Fermi length~\cite{Zunger1981,Lundqvist1976}. Such an approximation
works well inside the FM leads. However, inside the insulating layer the
electron densities are low and the screening length exceeds the layer thickness.
One can expect that the size of the exchange-correlation hole essentially
exceeds the thickness of the insulating layer. In this case the Coulomb interaction
is not the short range interaction any more. In this paper we take into account the
long-range nature of the Coulomb interaction inside the insulating layer.
Therefore, the LSD approximation can not be used to describe effects discussed in this paper.

5) The tunnelling matrix elements responsible
for spin currents and the hopping based IEC also depend on the dielectric
properties of the insulating matrix. In our model we assume that potential
profiles for electrons $U_{1,2}$ are the step functions. In reality the shape of potentials
is affected by the dielectric constant of the insulating layer (see \cite{Lundqvist}).
This effect requires a separate consideration.

\section{Conclusion}

We studied the exchange coupling in MTJ consisting of two FM layers separated by the insulating layer.
We calculated the exchange coupling due to many-body effects (the inter-electron Coulomb interaction).
The basic idea behind this mechanism is related to the fact that
the wave functions of electrons located at different FM
leads are overlapped inside the insulating layer.
In combination with weak screening of electric field inside the insulator these electrons
experience the indirect Coulomb interaction leading to interlayer magnetic coupling.
We showed that the Coulomb based IEC can exceed the hopping based exchange contribution
and found that these two mechanisms have essentially different dependence on system parameters.

We found that, in contrast to hopping based exchange, the Coulomb based exchange
interaction is inversely proportional to the dielectric constant $\varepsilon$ of the insulating layer.
The dependence of the IEC on the dielectric properties of the insulating layer in MTJ is similar
to ME effect where electric and magnetic degrees of freedom are coupled.

We calculated the IEC as a function of temperature and electric field (or voltage across the MTJ)
for MTJ with insulating layer made of TTF-CA and Hf$_{0.5}$Zr$_{0.5}$O$_2$. We showed that
in the vicinity of the FE phase transition of TTF-CA the IEC experiences large
variations. We found that even the FM-AFM transition may occur in MTJ
for some system parameters.

We found that IEC as a function of electric field (or applied voltage)
shows strong variations. The electric field
can cause the transition from AFM to FM coupling: for zero field the coupling is AFM
while for finite field it is FM.
This effect demonstrates the ME coupling in MTJ.

\ack

This research was supported by NSF under Cooperative Agreement Award EEC-1160504,
the U.S. Civilian Research and Development Foundation (CRDF Global)  and NSF PREM Award.
O.U. was supported by Russian Science Foundation (Grant  16-12-10340).

\appendix 

\section{Contribution of tunnelling to the Coulomb based IEC}\label{App:Validity}

The zero order wave function for FM configuration (the AFM wave function
can be considered in a similar way) is given by the following Slater determinant
\begin{equation}\label{Eq:ZeroOrderWF}
\Psi_0=\frac{1}{\sqrt{N}}\left(\begin{array} {ccc} \psi^{s_1}_1(\bi r_1) & \ldots & \psi^{s_1}_1(\bi r_{2n_0})\\ \vdots
&\ddots &\vdots\\
\phi^{s_{n_0+1}}_1(\bi r_1)& \ldots &\phi^{s_{n_0+1}}_1(\bi r_{2n_0})\\
 \vdots
&\ddots &\vdots\end{array}\right).
\end{equation}
States $\psi_i$ and $\phi_j$ are chosen such that all the
energy levels below $E_\mathrm F$ are filled: $n_0$ states in the
left lead and $n_0$ states in the right lead. $N$ is the normalization
factor. Further we introduce the excited states as follows
\begin{equation}\label{Eq:ExcWF}
\begin{array}{l} \Psi_{ij}^s=\\~\\=\hat b^{s+}_i \hat a^s_j \Psi_0 \\ ~ \\~ \end{array}= \frac{1}{\sqrt{N_{ij}}}\left(\begin{array}{ccc} \psi^{s_1}_1(\bi r_1) & \ldots & \psi^{s_1}(\bi r_{2n_0})\\
\vdots
&\ddots &\vdots \\
\phi^{s_{n_0+j-1}}_{j-1}(\bi r_1)&\ldots &\phi^{s_{n_0+j-1}}_{j-1}(\bi r_{2n_0})\\
\psi^{s}_{i}(\bi r_1)&\ldots &\psi^{s}_{i}(\bi r_{2n_0})\\
\phi^{s_{n_0+j+1}}_{j+1}(\bi r_1)&\ldots &\phi^{s_{n_0+j+1}}_{j+1}(\bi r_{2n_0})\\
\vdots
&\ddots &\vdots\end{array}\right).
\end{equation}
The annihilation operator removes a line in the Slater determinant
while the creation operator adds a line. $N_{ij}$ is the normalization factor.
We introduce the excited wave function as $\tilde \Psi^s_{ij}=\hat a^{s+}_i \hat b^s_j \Psi_0$.
These wave functions correspond to single excitations with only one electron transferred between
leads.  Using the above excited states we can write the
perturbed wave function as follows (see~\cite{Bel2016ExGr})
\begin{equation}
\Psi=(1+\alpha_0)\Psi_0+\!\!\!\sum_{s,i\notin S_0^s,j\in S_0^s} \beta_{ij}^s\Psi^s_{ij}+\!\!\!\sum_{s,i\notin S_0^s,j\in S_0^s} \tilde\beta_{ij}^s\tilde\Psi^s_{ij},
\end{equation}
with
\begin{equation}\label{Eq:WaveFuncFM1}
\beta_{ij}^s=\tilde\beta_{ij}^s=-\frac{T^s_{ij}}{\epsilon_i^s-\epsilon_j^s+ \epsilon_\mathrm c},
\end{equation}
\begin{equation}\label{Eq:PartNumbFM}
\alpha_0=-\!\!\!\!\!\sum_{s,i\notin S_0^s,j\in S_0^s}\!\!\frac{|  T^s_{ij}|^2}{(\epsilon_i^s-\epsilon_j^s+\epsilon_\mathrm c)^2},
\end{equation}
where $T_{ij}^s$ is the tunnelling matrix element, $\epsilon_\mathrm c$ is the charging energy, $S_0$ is the set of states in a single lead below the Fermi energy. The charging energy tends to zero for infinite leads, but it should still exceed the mean level spacing $\Delta$ ($\epsilon_\mathrm c\sim 1/\Omega^{1/3}$, $\Delta\sim 1/\Omega$). Note that excited states with two or more electrons transferred from one lead to another are also possible.
These states can be considered in a similar way. The Coulomb contribution to the energy of FM configuration is given by
\begin{equation}\label{Eq:EnCoulTot}
\eqalign{
H^\mathrm{FM}_\mathrm C=\langle\Psi|\hat H_\mathrm C|\Psi\rangle=\cr
=(1+2\alpha_0)\langle\Psi_0|\hat H_\mathrm C|\Psi_0\rangle+2\sum_{ij}|\beta_{ij}^s|^2\langle\Psi_{ij}|\hat H_\mathrm C|\Psi_{ij}\rangle+....
}
\end{equation}
In our manuscript we change it with the expression $\langle\Psi_0|\hat H_\mathrm C|\Psi_0\rangle$. The terms of the type $\langle\Psi_0|\hat H_\mathrm C|\Psi_{ij}\rangle$ not written in  ~(\ref{Eq:EnCoulTot}) correspond to tunnelling of electrons between the leads due to the Hamiltonian $\hat H_\mathrm C$.
These terms in fact are taken into account in tunnelling matrix elements $T^s_{ij}$.
The potential profile for s-electrons, $\hat U_{1,2}$ appears due to the mean electrostatic
potential created by ions and s-electrons themselves. Also, we neglect the
electron-electron scattering within leads, $\langle\Psi_{lk}|\hat H_\mathrm C|\Psi_{ij}\rangle$ ($k\ne i,j$, $l\ne i,j$).

The coefficients $\beta^s_{ij}\sim \rme^{-\varkappa d}$ are small. Therefore, in the last term
of  ~(\ref{Eq:EnCoulTot}) we can calculate $\langle\Psi_{ij}|\hat H_\mathrm C|\Psi_{ij}\rangle$ considering
leads as independent (non-interacting). The state $\Psi_{ij}$ is different from the state $\Psi_{0}$
by adding one electron into one lead and removing of one electron from
another lead. The Coulomb energy of a metallic lead is given by the expression~\cite{Glazman2002}
\begin{equation}\label{Eq:EnCoulLead}
U^\mathrm{L}_\mathrm C=\epsilon_\mathrm c n^2+\frac{\Delta}{\pi}(S-S_0)^2,
\end{equation}
where $n$ is the excessive charge and $S-S_0$ is the excessive spin (deviation of the
total spin from its equilibrium value). The non-zero equilibrium spin, $S_0$ appears in
our consideration due to the interaction with d-electrons. Taking  ~(\ref{Eq:EnCoulLead}) into account we can write
\begin{equation}\label{Eq:EnCoulLead1}
\langle\Psi_{ij}|\hat H_\mathrm C|\Psi_{ij}\rangle\approx \langle\Psi_{0}|\hat H_\mathrm C|\Psi_{0}\rangle+\epsilon_\mathrm c+O(\Delta).
\end{equation}
Substituting  ~(\ref{Eq:PartNumbFM}) and (\ref{Eq:EnCoulLead1}) into  ~(\ref{Eq:EnCoulTot}) we get
\begin{equation}\label{Eq:EnCoulTot1}
H^\mathrm{ex}_\mathrm C=\langle\Psi_0|\hat H_\mathrm C|\Psi_0\rangle+4\epsilon_c\sum_{ij}|\beta_{ij}^s|^2+....
\end{equation}
The hopping based contribution can be written using the same language in the form
\begin{equation}\label{Eq:EnHopTot1}
H^\mathrm{FM}_0=...+\sum_{ij}|\beta_{ij}^s|^2\langle\Psi_{ij}|\hat H_0|\Psi_{ij}\rangle,
\end{equation}
where
\begin{equation}\label{Eq:EnHop}
\langle\Psi_{ij}|\hat H_0|\Psi_{ij}\rangle=\epsilon_i^s-\epsilon_j^s+\langle\Psi_{0}|\hat H_0|\Psi_{0}\rangle.
\end{equation}
 Where $\epsilon_i-\epsilon_j$ is the difference of kinetic energies in states $i$ and $j$. This term describes virtual transitions of electrons from state $i$ (under the Fermi surface) in one lead to state $j$ (above the Fermi level) in another lead. Effective energy region in the vicinity of the Fermi level contributing to the hopping based exchange interaction can be estimated as follows $\hbar^2 \kappa_0/d\approx 100$ K (see~\cite{Slonczewski1989}). We can estimate $\epsilon_i-\epsilon_j\sim \hbar^2 \kappa_0/d\approx 100$ K. Using this estimation we have
 \begin{equation}\label{Eq:EnHop2}
H^\mathrm{FM}_0=...+\frac{\hbar^2 \kappa_0}{d}\sum_{ij}|\beta_{ij}^s|^2.
\end{equation}
Obviously, $\hbar^2 \kappa_0/d\gg \epsilon_\mathrm c$.
Comparing  ~(\ref{Eq:EnHop2}) and (\ref{Eq:EnCoulTot1}) one can see that
the last term in  ~(\ref{Eq:EnCoulTot1}) can be neglected
in comparison to the hopping based contribution.

Note that the summation in the matrix element
$\langle\Psi_0|\hat H_\mathrm C|\Psi_0\rangle$ is over all pair of electrons since
the Coulomb matrix element does not depend on the perpendicular momentum. The
matrix element $\beta^s_{ij}$ is non-zero only for transitions with conservation
of perpendicular momentum $\bi k_\perp$. Therefore, summation in the last term of
 ~(\ref{Eq:EnCoulTot1}) is over pairs of states with the same $\bi k_\perp$. Therefore,
there are much less pairs participating in the last term of  ~(\ref{Eq:EnCoulTot1}) in
comparison to the number of state pairs contributing to the first term of
 ~(\ref{Eq:EnCoulTot1}), and $\langle\Psi_0|\hat H_\mathrm C|\Psi_0\rangle\gg\epsilon_c\sum_{ij}|\beta_{ij}^s|^2$.
Thus, we can neglect the last term in  ~(\ref{Eq:EnCoulTot1}) and calculate the Coulomb
based IEC averaging over the unperturbed function $\Psi_0$.

\section{Contribution to the IEC from the insulating layer}\label{App:Calc}

Here we calculate the following integral
\begin{equation}\label{Eq:ExEnEst3}
\eqalign{
\tilde I^-_\mathrm{ex}= \frac{-\Omega^{2}}{(2\pi)^6S}\int \rmd^3 r_1 \rmd^3 r_2 \rmd^3k_{1} \rmd^3k_{2}\hat U^\mathrm I_\mathrm C(\bi r_1,\bi r_2) \times\cr \times \tilde\rho(z_1)(\tilde\rho(z_2))^*\rme^{\rmi(\bi k_{1\perp}-\bi k_{2\perp})\bi r_{1\perp}}\rme^{-\rmi(\bi k_{1\perp}-\bi k_{2\perp})\bi r_{2\perp}}+.
}
\end{equation}
The integration over the coordinates $\bi r_1$ and $\bi r_2$ in  ~(\ref{Eq:ExEnEst3})
is performed over the region between the leads. Renaming variables $\bi k_{1\perp}\to\bi k_{2\perp}$ and $\bi k_{2\perp}\to\bi k_{1\perp}$ makes
the integrand complex conjugate, however the whole integral stays the same meaning
that only the real part of the integral is finite. Below we consider the real part of the integrand
\begin{equation}\label{Eq:ExEnEst41}
\eqalign{
\tilde I^-_\mathrm{ex}= \frac{-\Omega^{2}}{(2\pi)^6S}\int \rmd^3 r_1 \rmd^3 r_2 \rmd^3k_{1} \rmd^3k_{3}\times\cr \times \tilde\rho(z_1)\hat U^\mathrm I_\mathrm C(\bi r_1,\bi r_2)(\tilde\rho(z_2))^*\times \cr  \{\cos((\bi k_{1\perp}-\bi k_{2\perp})\bi r_{1})\cos((\bi k_{1\perp}-\bi k_{2\perp})\bi r_{2})+\cr +\sin((\bi k_{1\perp}-\bi k_{2\perp})\bi r_{1})\sin((\bi k_{1\perp}-\bi k_{2\perp})\bi r_{2})\}.
}
\end{equation}
Both terms (with $\sin()$ and $\cos()$) give the same contribution. Thus, we consider one term with numerical
factor~2
\begin{equation}\label{Eq:ExEnEst5}
\eqalign{
\tilde I^-_\mathrm{ex}= \frac{-2\Omega^{2}}{(2\pi)^6S}\int \rmd^3 r_1 \rmd^3 r_2 \rmd^3k_{1} \rmd^3k_{2} \times\cr \times \tilde\rho(z_1)\hat U^\mathrm I_\mathrm C(\bi r_1,\bi r_2)(\tilde\rho(z_2))^*\times \cr \times \cos((\bi k_{1\perp}-\bi k_{2\perp})\bi r_{1})\cos((\bi k_{1\perp}-\bi k_{2\perp})\bi r_{2}).
}
\end{equation}
The last integral can be written in the form
\begin{equation}\label{Eq:ExEnEst6}
\eqalign{
\tilde I^-_\mathrm{ex}=\frac{-4\Omega^{2}}{(2\pi)^6S} \int_{0}^{k^-_\mathrm F}\!\!\! \rmd k_{1z} \int_{0}^{k_{1z}}\!\!\!\rmd k_{2z}\int \rmd^3 r_1 \rmd^3 r_2 \times\cr \times\int_{|\bi k_{1\perp}|^2<(k_\mathrm F^-)^2-k^2_{1z}}\!\!\!\!\!\!\!\!\!\!\!\!\!\rmd^2k_{1\perp}\int_{|\bi k_{2\perp}|^2<(k_\mathrm F^-)^2-k^2_{2z}}\!\!\!\!\!\!\!\!\!\!\!\!\! \rmd^2k_{2\perp}\tilde\rho(z_1)(\tilde\rho(z_2))^*\times \cr \times \hat U^\mathrm I_\mathrm C(\bi r_1,\bi r_2)\cos((\bi k_{1\perp}-\bi k_{2\perp})\bi r_{1})\cos((\bi k_{1\perp}-\bi k_{2\perp})\bi r_{2}).
}
\end{equation}
Equation~(\ref{Eq:ExEnEst6}) shows that the exchange interaction is the sum of matrix
elements of the Coulomb interaction between the effective charges. These charges decay
toward the middle of the barrier, $z=0$, and oscillate in the $(x,y)$-plane.
The oscillations are coherent and have the same phase.

The normalized matrix element of the effective Coulomb interaction is given by the expression
\begin{equation}\label{Eq:CoulIntME}
\eqalign{
W_\mathrm I=\frac{1}{S}\int \rmd^3 r_1 \rmd^3 r_2 \tilde\rho(z_1)(\tilde\rho(z_2))^*\times \cr \times \hat U^\mathrm I_\mathrm C(\bi r_1,\bi r_2)\cos((\bi k_{1\perp}-\bi k_{2\perp})\bi r_{1})\cos((\bi k_{1\perp}-\bi k_{2\perp})\bi r_{2}).
}
\end{equation}
We introduce the notations $\rho_0^2=|e\tau_{i}^{s*}\tau^{s'}_{j}/\Omega|^2$, $\Delta \varkappa = \varkappa_1-\varkappa_2$, $\bi q=\bi k_{1\perp}-\bi k_{2\perp}$ ($q=|\bi q|$).  To evaluate the matrix element we calculate the
electric field created by the effective charges. Due to periodic harmonic
variations of the electric charge the field has the same harmonic variations.
The electric field inside the metallic leads is zero since it is screened
by the surface charges which vary with the wave vector $\bi q$.

Consider the z-component of the electric field at $\bi r_{\perp}=0$.
At this symmetry point the field has no (x,y)-plane components. The field created by the charge in the gap has the form
\begin{equation}\label{Eq:Field1}
\eqalign{
E^{\mathrm{ch}}_{z}(z)=\frac{\rho_0}{4\pi\varepsilon_0\varepsilon}\int_0^{2\pi}\!\!\! \rmd\varphi \int_0^{\infty}\!\!\! r_\perp \rmd r_\perp \int_{-d/2}^{d/2}\!\!\! \rmd\tilde z \times \cr \times \frac{ \cos(qr_\perp \cos(\varphi))(z-\tilde z)\rme^{-\Delta \varkappa \tilde z}}{(r_\perp^2+(z-\tilde z)^2)^{3/2}}=\cr =\frac{\rho_0}{2\varepsilon_0\varepsilon}\left\{\frac{\rme^{-qz}}{q-\Delta \varkappa}(\rme^{(q-\Delta \varkappa)z}-\rme^{-(q-\Delta\varkappa)d/2})+\right.\cr +\left.\frac{\rme^{qz}}{q+\Delta \varkappa}(\rme^{-(q+\Delta \varkappa)d/2}-\rme^{-(q+\Delta\varkappa)z})\right\}.
}
\end{equation}
The electric field created by the surface metal charges is given by the expression
\begin{equation}\label{Eq:Field2}
E^{\mathrm{m}}_{z}(z)=\frac{\rho_0}{2\varepsilon_0\varepsilon}(\sigma_2\rme^{-q(z+d/2)}+\sigma_1\rme^{q(z-d/2)}),
\end{equation}
where the surface charges $\sigma_{1,2}$ are
\begin{equation}\label{Eq:SurfCharges}
\sigma_{1(2)}=\frac{\sigma_{1(2)}^0\rme^{qd}+\sigma_{2(1)}^0}{\rme^{qd}-\rme^{-qd}},
\end{equation}
with
\begin{equation}\label{Eq:SurfCharges2}
\eqalign{
\sigma_1^0=\frac{\rme^{-qd/2}}{q-\Delta \varkappa}\left(\rme^{(q-\Delta\varkappa)d/2}-\rme^{-(q-\Delta\varkappa)d/2}\right),\cr
\sigma_2^0=\frac{\rme^{-qd/2}}{q+\Delta \varkappa}\left(\rme^{-(q+\Delta\varkappa)d/2}-\rme^{(q+\Delta\varkappa)d/2}\right).
}
\end{equation}
The total value of the z-component of electric field due to
the charge $\rho^{(1)}$ is the sum, $E^{(1)}_z=E^{\mathrm{ch}}_{z}+E^{\mathrm{m}}_{z}$.
Similarly we can calculate the amplitude of the in-plane field component $E^{(1)}_x$.
Its spatial variation is shifted by a half period $1/(2q)$.

Both charges in  ~(\ref{Eq:ExEnEst6}) have the same period of oscillations
and the same phase. The field produced by the charge $\rho^{(2)}$ is the same as the field
$\bi E^{(1)}$, $(E^{(2)}_z(z)=E^{(1)}_z(z))$, $(E^{(2)}_x(z)=E^{(1)}_x(z))$.
The surface density (density per surface area of leads) of interaction energy can be calculated as follows
$W_\mathrm I=(\varepsilon_0\varepsilon/2)(\int dz E^{(1)}_z(z) E^{(1)}_z(z)+\int dz E^{(1)}_x(z) E^{(1)}_x(z))$, where
integration goes over the region $|z|<d/2$. Finally, the matrix element can be estimated as follows
\begin{equation}\label{Eq:IntEnTot}
W_\mathrm I=W_{\mathrm Ix}+W_{\mathrm Iz},
\end{equation}
where
\begin{equation}\label{Eq:IntEn}
\eqalign{
W_{\mathrm Iz}=\frac{\rho_0^2}{8\varepsilon_0\varepsilon}\left\{(\alpha_1^2+\alpha^2_2)\frac{\mathrm{sinh}(dq)}{q}+\alpha_3^2\frac{\mathrm{sinh}(d\Delta\varkappa)}{q}\right.\cr +2\alpha_1\alpha_2d+4\alpha_1\alpha_3\frac{\mathrm{sinh}(\!(\Delta\varkappa+q)d/2)}{\Delta\varkappa+q}\!\cr 
\left.+4\alpha_2\alpha_3\frac{\mathrm{sinh}(\!(\Delta\varkappa-q)d/2)}{\Delta\varkappa-q}\!\!\right\},\cr
W_\mathrm {Ix}=\frac{\rho_0^2}{8\varepsilon_0\varepsilon}\left\{(\tilde\alpha_1^2+\tilde\alpha^2_2)\frac{\mathrm{sinh}(dq)}{q}+\tilde\alpha_3^2\frac{\mathrm{sinh}(d\Delta\varkappa)}{q}\right.\cr +2\tilde\alpha_1\tilde\alpha_2d+4\tilde\alpha_1\tilde\alpha_3\frac{\mathrm{sinh}(\!(\Delta\varkappa+q)d/2)}{\Delta\varkappa+q}\!\cr 
\left.+4\tilde\alpha_2\tilde\alpha_3\frac{\mathrm{sinh}(\!(\Delta\varkappa-q)d/2)}{\Delta\varkappa-q}\!\!\right\},
}
\end{equation}
with
\begin{equation}\label{Eq:IntEn21}
\alpha_1=\rme^{-qd/2}\sigma_2-\frac{\rme^{-(q-\Delta\varkappa)d/2}}{q-\Delta\varkappa},
\end{equation}
\begin{equation}\label{Eq:IntEn22}
\tilde \alpha_1=-\rme^{-qd/2}\sigma_2-\frac{\rme^{-(q-\Delta\varkappa)d/2}}{q-\Delta\varkappa},
\end{equation}
\begin{equation}\label{Eq:IntEn23}
\alpha_2=\rme^{-qd/2}\sigma_1+\frac{\rme^{-(q+\Delta\varkappa)d/2}}{q+\Delta\varkappa},
\end{equation}
\begin{equation}\label{Eq:IntEn24}
\tilde \alpha_2=\rme^{-qd/2}\sigma_1-\frac{\rme^{-(q+\Delta\varkappa)d/2}}{q+\Delta\varkappa},
\end{equation}
\begin{equation}\label{Eq:IntEn25}
\alpha_3=\frac{2\Delta\varkappa}{q^2-\Delta\varkappa^2}, ~~\tilde \alpha_3=\frac{-2q}{q^2-\Delta\varkappa^2}.
\end{equation}
Introducing $W_\mathrm I$ into  ~(\ref{Eq:ExEnEst6}) we arrive to the final expression for $\tilde I^-_\mathrm{ex}$
\begin{equation}\label{Eq:ExEnEst83}
\eqalign{
\tilde I^-_\mathrm{ex}= -\frac{e^2}{32\pi^4\varepsilon_0\varepsilon} \int_{0}^{k^-_\mathrm F}\int_{0}^{k_1}\!\!\! \rmd k_{1}\rmd k_{2} |(\tau^-_1)^*\tau^-_2|^2 \rme^{-d(\varkappa^-_1+\varkappa^-_2)} \times \cr \times \int_{0}^{k_2^{\mathrm{max}}+k_1^{\mathrm{max}}}q \omega_\mathrm I (q) \rmd q \!\!\int_0^{(k_2^{\mathrm{max}}+k_1^{\mathrm{max}})/2}\!\!\! k\zeta(k,q)\rmd k,
}
\end{equation}
where
\begin{equation}\label{Eq:InPlaneInt6}
\omega_\mathrm I (q) =\frac{8\varepsilon_0\varepsilon W_\mathrm I(q)}{\rho_0^2}.
\end{equation}
We introduced the following functions
\begin{equation}\label{Eq:InPlaneInt2}
\eqalign{
\zeta=\left\{\begin{array}{l}0,~~(\phi_2<\phi_3)~ \mathrm{or}~ (\phi_1<\phi_3),\cr
\phi_1-\phi_3,~~\mathrm{otherwise,}\end{array}\right.
}
\end{equation}
where
\begin{equation}\label{Eq:InPlaneInt3}
\eqalign{
\phi_1(k,q)=\left\{\begin{array}{l}0,~~k>k_1^{\mathrm{max}}+q/2,\cr
\frac{\pi+\pi\mathrm{sign}(k_1^{\mathrm{max}}-q/2)}{2},~~k<|k_1^{\mathrm{max}}-q/2|,\cr
\mathrm{arccos}\left(\frac{k^2+q^2/4-(k_1^{\mathrm{max}})^2}{qk}\right),~~\mathrm{otherwise}. \end{array}\right.
}
\end{equation}
\begin{equation}\label{Eq:InPlaneInt4}
\eqalign{
\phi_2(k,q)=\left\{\begin{array}{l}\pi,~~k<k_2^{\mathrm{max}}-q/2,\cr
\mathrm{arccos}\left(\frac{k^2+q^2/4-(k_2^{\mathrm{max}})^2}{qk}\right),~~\mathrm{otherwise}. \end{array}\right.
}
\end{equation}
\begin{equation}\label{Eq:InPlaneInt5}
\phi_3(k,q)=\pi-\phi_2(k,q).
\end{equation}
In a similar way one can calculate the integrals $\tilde I^+_\mathrm{ex}$ and $\tilde{\tilde I}_\mathrm{ex}$.

\section*{References}

\bibliography{Exchange}

\end{document}